\documentclass[11pt]{article}
\usepackage{latexsym,amssymb,amsmath}
\usepackage[dvips]{graphics,graphicx,epsfig,color}
\usepackage[lofdepth,lotdepth]{subfig}
\begin{document}
%
\thispagestyle{empty}
\title{\bf  Incorporation of macroscopic heterogeneity within a porous layer to enhance its acoustic absorptance}
\author{\bf Armand Wirgin \thanks {
LMA, CNRS, UMR 7031, Aix-Marseille Univ, Centrale Marseille, F-13453 Marseille Cedex 13, France.}}
\date{\today}
\maketitle
\begin{abstract}
We seek the response, in particular the spectral absorptance, of a rigidly-backed periodically-(in one horizontal~~ direction) ~inhomogeneous ~layer ~composed ~of ~alternating rigid and macroscopically-homogeneous porous portions,   submitted to an airborne acoustic plane body wave. The rigorous theory of this  problem is given and the means by which the latter can be numerically solved are outlined. At low frequencies, a suitable approximation derives from one linear equation in one unknown. This approximate solution is shown to be equivalent to that of the problem of the same  wave incident on a  homogeneous, isotropic layer. The thickness $h$ of this layer is identical  to that of the inhomogeneous layer, the effective complex body wave velocity therein is identical to that of the porous portion of the inhomogeneous layer, but the complex effective mass density, whose expression is given in explicit algebraic form, is  that of the reference homogeneous macroscopically-porous layer divided by the filling factor (fraction of porous material to the total material in one grating period). This difference of density is the reason why it is possible for the lowest-frequency absorptance peak to be higher than that of a reference layer.  Also, it is shown how to augment the height of this peak so that it attains unity (i.e., total absorption) and how to shift it to lower frequencies, as is required in certain applications.
\end{abstract}
Keywords: dynamic response, absorptance, inhomogeneous medium, porous component, grating, effective medium, .
\newline
\newline
Abbreviated title: An effective layer that absorbs like a grating when submitted to sound
\newline
\newline
Corresponding author: Armand Wirgin, \\ e-mail: wirgin@lma.cnrs-mrs.fr
 \newpage
\tableofcontents
\newpage
\section{Introduction}\label{intro}
The absorption of waves (hydrodynamic waves, optical or other electromagnetic waves, acoustic waves in fluids, elastic waves in solids) is an ongoing major research topic. Four applicative subtopics illustrate this importance: i) harvesting of energy \cite{mm13}, particularly that coming from solar irradiation \cite{ta61,ho63,cz75,th75,pt75,go78,ho78,sc79,wi81,ng83,pm08,sv08,tg08,bg10,gh11,yg12,dl12,ny14,lc15,lgy16}, ii) breaking  (i.e., dissipating the energy of) water waves in rising oceans to protect harbors, coastal shores and constructions \cite{ll93,ep97,pe99,ue99,pe05,yg12,as14,zx15,nk16},  iii) sound level reduction in increasingly noisy open environments and reverberation reduction in enclosures \cite{db67,ak76,m80,an85,ac10,gdud11,nr12,wl12,hl12,gl15,gh15,gp16,lg16,hr17,at18,jrg18}, iv) mitigation of earthquake effects in nuclear power plants and cities \cite{gr05,th12,cb15,pk16,au16,au16,ww18,wi18b,wi18d}. For  all of these problems, solutions have been explored in which the principal dissipative agent is a natural or man-made  film, barrier,  layer, or solid object (e.g., plate) or medium (e.g., portion of the underground) of finite volume in (or on) which inclusions or voids are placed and arranged either in random or periodic manner. Naturally, the interaction of waves with these heterogeneous media has given rise to a considerable amount of theoretical/numerical research to try to predict and optimize the outcome of the interaction. The present contribution is of this sort and concentrates on the acoustic wave absorption problem.

  Deliberate research on the latter topic probably begins in 1907 when Rayleigh  \cite{ra07} elaborates the first mathematical (called by him 'dynamical') theory of the diffraction by periodically-uneven (in one direction) surfaces and interfaces (note that the region between the lowest and highest points of such a structure can be considered to be an inhomogeneous layer, comprised alternatively of the materials in the upper and lower media), called 'gratings', applicable at all frequencies and for all grating periods. After showing that the high frequency (Fresnel theory) approximation of grating response predicts that {\it none} of the incident energy is thrown into the zeroth (specular) reflected order for certain rigid and pressure release gratings, he shows that his own theory predicts, for gratings whose grooves (called by him 'corrugations') have any shape, and whatever be the nature (rigid or pressure release) of the grating material, that {\it all} of the incident energy is thrown into the specular reflected order when the grating period is inferior to the wavelength of the incident sound.  Rayleigh concludes that there is a need for more detailed examination of what actually goes on  near the grating surface (his previous finding applying only to the grating orders which are far-field entities) when its period is small relative to, or near, the sound wavelength $\lambda$.

 In another contribution \cite{ra45}, Rayleigh writes: "The above investigation (i.e., that in \cite{ra07}) is limited to the case where the second medium is  impenetrable, so that the whole energy of the incident wave is thrown back in the regularly reflected wave and in the diffracted spectra. It is an interesting question whether the conclusion that corrugations of period less than $\lambda$ have no effect can be extended so as to apply when there is a wave regularly transmitted. It is evident that the principle of energy does not suffice to decide the question, but it is probable that the answer should be in the negative. If we suppose the corrugations of given period to become very deep and involved, it would seem that the condition of things would at last approach that of a very gradual transition between the media, in which case  the reflection tends to vanish". As far as we know, Rayleigh's intuition (now known as the 'moth's eye principle' \cite{ta61,th75,lc15,cd18}) was neither given a theoretical justification by him, nor even exploited until the advent of  anechoic chambers incorporating walls and ceilings with deeply-corrugated linings covering  porous materials.

It has long been known (empirically) that many naturally- porous materials and media such as turf, dry sand, ashes, asbestos and especially snow \cite{ke40}, constitute excellent absorbers of audible sound. Porous man-made materials such as  carpets, fibre glass, and rubber-like or plastic foams also have this feature and have long been incorporated in living quarters, work spaces and concert halls to reduce the reverberation of, and thus dissipate,  audible sounds, be the latter agreeable or disagreeable (noise). Anechoic chambers (see \cite{em73,sa05,tb08,lh09,ck12,drt14,kd14} for electromagnetic microwave absorption materials, notably, but not exclusively, destined to anechoic chambers) incorporate rather thick linings of manufactured porous materials, but this is usually not sufficient to reduce the reverberations, notably in the low frequency range, to acceptable levels, whence the idea of superposing periodic arrays of absorbing objects such as pyramids on the porous lining so as to provoke reduced reflection (and increased absorption, hopefully for all incident angles and over substantial bandwidths) by what may be thought to be the moth's eye effect \cite{lc15}.

 Perhaps the first attempt at explaining the acoustic moth's eye effect is due to De Bruijn  \cite{db71} who carries out a rigorous analysis of the interaction of a plane acoustic body wave with a lamellar grating (periodic in one direction, grooves of rectangular shape), the boundary of which is composed of rigid vertical strips and absorbing (simulated by an impedance boundary condition) horizontal strips. The analysis resides on the Rayleigh plane wave expansion in the air half space (wherein propagates the incident  wave) and a modal representation of the pressure field in the grooves.  De Bruijn is able to solve for the scattering amplitudes and reflectance (and thus the absorptance via the principle of conservation of energy) even when the grating period and groove depths are of the order of the wavelength and shows that the absorptance depends considerably  not only on the angle of incidence  but also on the period and dimensions of the grooves. He concludes  that the whole matter of absorption afforded by this structure is completely governed by three physical phenomena: 1. an anomaly due to the emergence of a new spectral order at grazing angle (Wood anomaly \cite{ms16}, characterized by sudden changes in the reflectance as a function of wavelength and explained by Rayleigh as occurring when a body plane wave in the plane wave representation changes to an evanescent plane wave); 2. resonances related to the guided complex wave supportable by the grating (surface resonance); 3. resonances related to the depth of the groove (which is a sort of waveguide). This conclusion (the optical analogs of which are \cite{gv99,gm02,pm04,bt08,lp08}), especially the last one, constitutes an important first step in the comprehension of why corrugations can modify the absorption of otherwise flat  absorbing surfaces.

Bos et al.  \cite{bd05} continue the study of De Bruijn by replacing the 1D grating by a 2D (i.e., periodic in two orthogonal directions) grating with box-like grooves. This grating is rigid and placed over, and in contact, with a semi-infinite porous half space absorbing medium whose constitutive properties are described by the semi-empirical Delany and Blazey model \cite{db70}. Particularly interesting is the result depicted in their fig. 4 concerning the existence, for normal incidence airborne sound, of a total absorption peak (see \cite{mp76,hm76,mn77,cs80,st81,pt92,rb16} for the optical analogs thereof) at  a frequency slightly higher than $600~Hz$ which is far above the $~\sim 0.25$ absorptance obtained at this frequency without the presence of the grating.

Groby et al. \cite{gw09} tackle the problem of reflection, transmission and absorption of airborne sound by a porous layer containing a 1D periodic set of macroscopic circular cylindrical fluid-like inclusions. Their rigorous analysis appeals to the Rayleigh plane wave expansions as well as multipole expansions to account for the presence of the cylinders. They show that high-contrast inclusions in a porous plate induce an increase in the absorption coefficient, mainly associated with a decrease
in the hemispherical transmission coefficient for frequencies
that are higher than a frequency offset ($\sim 20~kHz$). They attribute this effect to the excitation of what they call 'modified plate modes', an explanation that is not in contradiction  with  one of the conclusions of De Bruijn for a different periodic structure.

Groby, Lauriks  and Vigran  \cite{glv10} examine the acoustic properties of a low resistivity porous layer backed by a rigid plate containing a 1D periodic array of rectangular irregularities, a structure that constitutes a generalization of the one treated by De Bruijn.
Numerical results deriving from the rigorous solution, as well as experimental results, show that such a structure can give rise to a total absorption peak at the frequency of the modified mode of the layer.

Groby, Dazel et al. \cite{gdd11} study the absorptive properties of a rigidly-backed porous layer containing  a periodic set of rigid circular inclusions in response to an airborne acoustic  plane body wave.  They obtain numerical results from their rigorous theoretical solution which show that this inhomogeneous layer gives rise to a quasi-total  absorption peak below the quarter-wavelength resonance
of the corresponding macroscopically-homogeneous porous layer (i.e., the one not containing  inclusions). This result is explained by the excitation of a complex trapped mode whose characteristics are similar to those of \cite{gw09}. This article constitutes the primary inspiration of the present investigation so that we terminate here our review of the literature on this subject. Naturally, much more research has been carried out since 2011 on the absorption of acoustic waves in macroscopically-homogeneous or inhomogeneous porous media and some of the associated articles are referenced in the first lines of this section.

As seen from this literature review, most recent studies of the absorption of airborne sound concern a porous host layer with fluid-filled or rigid inclusions. In the following,  the host layer can indifferently be considered as either rigid with porous inclusions or porous with rigid inclusions. We shall show, by means of a very simple theory, derived from, and verified by, a rigorous dynamical theory, that increased absorption (over that of the  rigidly-backed porous layer without rigid inclusions) can be optimized so as to be total, and, in any case, that  the cause of the lowest-frequency enhanced absorption peak is related to something akin to the quarter wavelength pseudo resonance.
\section{Response of the macroscopically-inhomogeneous layer to an airborne  plane-wave solicitation}\label{exact}
%
\subsection{Description of the configuration}\label{dc}
 Let $Oxyz$ be a cartesian coordinate system  with origin at $O$, A flat-faced   layer is located between the planes $z=0$ and $z=h$. The medium underneath (i.e. in $z<0$) the layer is air and the medium above (i.e., $z>h$) the layer is a perfectly-rigid solid. The medium  within the layer (henceforth called a grating) is periodic in the $x$ direction and invariant (and infinite) in the $y$ direction. The period of the grating is $d$ and in each period two contiguous blocks, both of of height $h$, are present, one of width (along $x$) $w$ filled with a porous medium such as foam, this block being called 'groove' from now on, and the other of width $d-w$ filled with the same rigid solid as in $z>h$. The grating is thus a macroscopically-inhomogeneous bi-phasic layer whose porous component will be treated as a macroscopically-homogeneous medium.

The acoustic wave sources are assumed to be located in  the region beneath the layer  and to be infinitely-distant from $z=0$ so that the  solicitation takes the form of a body (plane) pressure wave in the neighborhood of the layer. The incident  wavevector $\mathbf{k}^{i}$ is assumed to lie in the $x-z$ plane, i.e., $\mathbf{k}^{i}=(k_{x}^{i},k_{y}^{i},k_{z}^{i})=(k_{x}^{i},0,k_{z}^{i})$, which fact, together with the independence of the grating geometry and the constitutive properties of the media in presence with respect to $y$, means that the incident and total pressure wave fields in the various domains do not depend on $y$. The  wavevector $\mathbf{k}^{i}$ of the plane wave solicitation is of the form $\mathbf{k}^{i}=(k^{[0]}\sin\theta^{i},0,k^{[0]}\cos\theta^{i})$ wherein  $\theta^{i}$ is the angle of incidence (see fig. \ref{blockabsgrat}), and $k^{[l]}=\omega/c^{[l]}~;~l=0,1$, with $\omega=2\pi f$ the angular frequency, $f$ the frequency, and $c^{[l]}$ the body wave velocity  in  air (for $l=0$) and in the porous material (for $l=1$). More often than not, we shall assume $\theta^{i}=0^{\circ}$.

Consequently,  the to-be-considered problem is 2D  and can be examined in the sagittal $x-z$ plane.
\begin{figure}[ht]
\begin{center}
\includegraphics[width=0.65\textwidth]{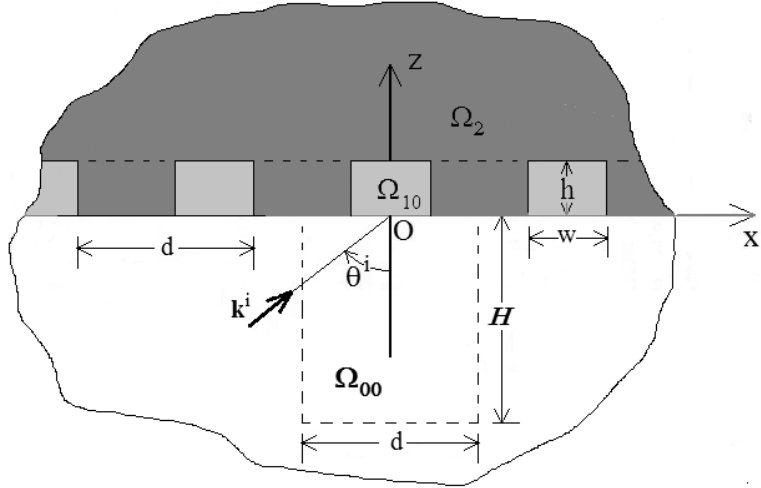}
\caption{Sagittal plane view of the problem. The absorbing structure can be viewed either as a rigid amellar grating whose grooves are entirely filled with a porous absorbing material or as a periodic  layer located between two flat planes $z=0$ and $z=h$   composed of alternating rigid and porous blocks (rectangles in the figure wherein the dashed line delineates the  upper 'boundary' of the layer). The half-space below $z=0$ ($\Omega_{0}$) is filled with air (white area in the figure). The  half space above $z=0$ not including the rectangular grooves is designated by $\Omega_{2}$ and is occupied by a rigid material (dark grey area in the figure).  The central layer domain is $\Omega_{10}$ (width $w$, height $h$), its left-hand neighbor is the domain $\Omega_{1-1}$ and its right-hand neighbor is the domain $\Omega_{11}$, etc. All the blocks of width $w$ are filled with an absorptive material ('foam'; light gray areas in the figure) whereas all the blocks of width $d-w$ are rigid. The grating, of period $d$, is solicited by an airborne acoustic plane body wave whose wavevector (lying in the sagittal plane) makes an angle $\theta^{i}$ with the $z$ axis.}
\label{blockabsgrating}
\end{center}
\end{figure}
Fig. \ref{blockabsgrat} depicts the problem  in the sagittal plane in which: $\Omega_{0}$ is what was previously below $z=0$, $\Omega_{2}$ is what was previously above $z>h$, and $\Omega_{1}=\cup_{n\in\mathbb{Z}}\Omega_{1n}$  is the layer-like (composite) domain constituted by the periodic assembly of blocks, with $\Omega_{1n}$ the $n$-th  block of rectangular cross section (width $w$ and height $h$).

The air medium in $z<0$ is assumed to be non-lossy and non-dispersive over the range of frequencies of interest, and its mass density  to be real. The (real) longitudinal-wave velocity in this fluid is the real constant  $c^{[0]}=\sqrt{\frac{K^{[0]}}{\rho^{[0]}}}$, with  $\rho^{[0]}$ the mass density and $K^{[0]}$ the isentropic bulk modulus.

As shown in sects. \ref{gdd} and \ref{glv}, the macroscopically-homogeneous foam medium in each groove  behaves like a lossy and dispersive fluid within which the  mass density $\rho^{[1]}(\omega)=\rho_{e}(\omega)$ and longitudinal wave velocity $c^{[1]}(\omega)=\sqrt{\frac{K^{[1]}}{\rho^{[1]}}}=c_{e}(\omega)$ are generally-complex functions of $\omega$.
\subsection{Effective-medium behavior of the macroscopically-homogeneous lossy filler material}\label{foam}
The theory of wave propagation in porous media, considered to be macroscopically-homogeneous, was initially elaborated by Biot \cite{bi56a,bi56b}. In most of the plastic foams saturated by a light fluid like air, the rigid frame assumption is valid so that an acoustic wave impinging on such a porous sample induces wave propagation only in the fluid phase. Therefore the viscothermal effects taking place in the pore channels are accounted-for by an effective density and an effective bulk modulus  of a so-called equivalent fluid \cite{bi56a,bi56b}. The  rigid frame model was extended to macroscopically-inhomogeneous porous media in \cite{jkd87,aa09,ca91}.

We shall apply what has become to be known as the Johnson-Champoux-Allard (JCA) model  to account for the absorption of airborne sound in the foam material component (considered here to be macroscopically-homogeneous) of our grating configuration.

In the frequency domain (the $\exp(-i\omega t$ temporal factor is implicit) the wave equation in terms of the fluid pressure $u$ inside the equivalent macroscopically-homogeneous or -inhomogeneous fluid is
\begin{equation}\label{0-000}
\nabla\left(\frac{\nabla u}{\rho_{e}(\omega)}\right)+\frac{\omega^{2}}{K_{e}(\omega)}u=0~.
\end{equation}
 Attenuation, viscothermal losses and dispersion are accounted-for in the complex effective density $\rho_{e}$ and effective bulk modulus $K_{e}$. The effective sound speed and characteristic impedance are $c_{e}(\omega)=\sqrt{K_{e}(\omega)/\rho_{e}(\omega)}$  and $Z_{e}(\omega)=\rho_{e}(\omega)c_{e}(\omega)$ \cite{aa09,dr07}.  The JCA expressions for $\rho_{e}$ and $K_{e}$ are:
\begin{equation}\label{0-010}
\rho_{e}(\omega)=\frac{\rho_{f}\tau_{\infty}}{\phi}\left(1+i\frac{\omega_{c}}{\omega}F(\omega)\right)~,
\end{equation}
\begin{equation}\label{0-020}
K_{e}(\omega)=\frac{\gamma P_{0}\phi^{-1}}
{
\gamma-(\gamma-1)
\left(
1+i
\frac{\omega^{'}_{c}}{\omega}
\frac{G(\omega)}{Pr}
\right)^{-1}
}~,
\end{equation}
wherein: $\omega_{c}=\sigma\phi/\rho_{f}\tau_{\infty}$ is the Biot frequency, $\omega'_{c}=\sigma'\phi/\rho_{f}\tau_{\infty}$ , $\gamma$ the specific heat ratio, $P_{0}$ the atmospheric pressure,  $Pr$ the Prandtl number (equal to 0.707 in  air at $25^{\circ}C$), $\rho_{f}$ the mass density of the fluid (in the interconnected) pores, $\phi$ the open-cell porosity, $\tau_{\infty}$ the high-frequency limit of tortuosity, $\sigma$ the static air flow resistivity, and $\sigma^{'}$ the static thermal resistivity. The correction functions $F$ and $G$ are
\begin{equation}\label{0-030}
F(\omega)=\sqrt{1-i\eta\rho_{f}\omega\left(\frac{2\tau_{\infty}}{\sigma\phi\Lambda}\right)^{2}}~,
\end{equation}
\begin{equation}\label{0-040}
G(\omega)=\sqrt{1-i\eta\rho_{f}Pr\omega\left(\frac{2\tau_{\infty}}{\sigma^{'}\phi\Lambda^{'}}\right)^{2}}~,
\end{equation}
in which $\eta$ is the dynamic viscosity of the fluid (here air), $\Lambda^{'}$ the thermal characteristic length of Champoux and Allard \cite{ca91}  and $\Lambda$ the viscous characteristic length of Johnson et al. \cite{jkd87}. The 'static' thermal resistivity is related to the thermal characteristic length via $\sigma^{'}=8\tau_{\infty}\eta/\phi\Lambda^{'2}$.

Note that the Biot characteristic angular frequency $\omega_{c}$ separates the low and high
frequency regimes (the viscous flow and inertial flow in the
pores), i.e.,  when $\omega < \omega_{c}$ , the viscous forces dominate and when $\omega > \omega_{c}$,
the inertial forces dominate.

Note also \cite{jkd87} that $\Re c_{e}(\omega)<\lim_{\omega\rightarrow\infty}\Re c_{e}(\omega)$ and \cite{jkd87}
\begin{equation}\label{0-045}
\lim_{\omega\rightarrow\infty} \Re c_{e}(\omega)=\frac{c_{f}}{\sqrt{\tau_{\infty}}}~,
\end{equation}
and since $\tau_{\infty}>1$, the effective phase velocity in the porous medium is less than the phase velocity in air for all finite frequencies. This essential feature of our porous media will be illustrated hereafter in figs. \ref{gdd11params} and \ref{glv10params}.
\subsubsection{The foam parameters in \cite{gdd11} and the associated effective density and wavespeed}\label{gdd}
The absorbing material employed in \cite{gdd11} is called Fireflex 2 which is an open-celled melamine foam. Its parameters are:\\
$\phi=0.95$\\
$\tau_{\infty}=1.42$\\
$\Lambda=180\times 10^{-6}~m$\\
$\Lambda'=360\times 10^{-6}~m$\\
$\sigma=8900~Pa~sm^{-2}$\\
$f_{c}=781~Hz$\\
$\rho_{f}=1.213~Kgm^{-3}$\\
$P_{0}=1.01325\times 10^{5}~Pa$\\
$\gamma=1.4$\\
$\eta=1.839\times 10^{-5}~Kgm^{-3}s^{-1}$.\\\\
With the above parameters, the velocity of sound in the pore fluid (here air) is
\begin{equation}\label{0-000}
    c_{f}=\sqrt{\frac{\gamma P_{0}}{\rho_{f}}}=341.973~ms^{-1}~.
\end{equation}

\begin{figure}[ht]
\begin{center}
\includegraphics[width=0.65\textwidth]{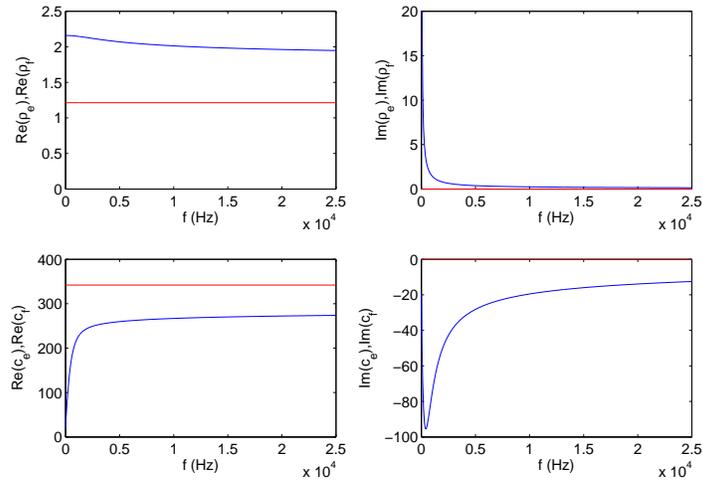}
\caption{Top left-hand panel is relative to $\Re \rho_{e}(f)$ (blue curve) and $\Re \rho_{f}(f)$ (red curve). Top right-hand panel is relative to $\Im \rho_{e}(f)$ (blue curve) and $\Im \rho_{f}(f)$ (red curve). Bottom left-hand panel is relative to $\Re c_{e}(f)$ (blue curve) and $\Re c_{f}(f)$ (red curve). Bottom right-hand panel is relative to $\Im c_{e}(f)$ (blue curve) and $\Im c_{f}(f)$ (red curve).}
\label{gdd11params}
\end{center}
\end{figure}

The evolution of the JCA effective constitutive parameters with frequency $f$ is given in fig. \ref{gdd11params}.
\clearpage
\newpage
\subsubsection{The foam parameters in \cite{glv10} and the associated effective density and wavespeed}\label{glv}
The absorbing material employed in \cite{glv10} is supposedly another  open-celled  foam. Its parameters are:\\
$\phi=0.96$\\
$\tau_{\infty}=1.07$\\
$\Lambda=273\times 10^{-6}~m$\\
$\Lambda'=672\times 10^{-6}~m$\\
$\sigma=2843~Pa~sm^{-2}$\\
$f_{c}=334~Hz$\\
$\rho_{f}=1.213~Kgm^{-3}$\\
$P_{0}=1.01325\times 10^{5}~Pa$\\
$\gamma=1.4$\\
$\eta=1.839\times 10^{-5}~Kgm^{-3}s^{-1}$.\\

\begin{figure}[ht]
\begin{center}
\includegraphics[width=0.65\textwidth]{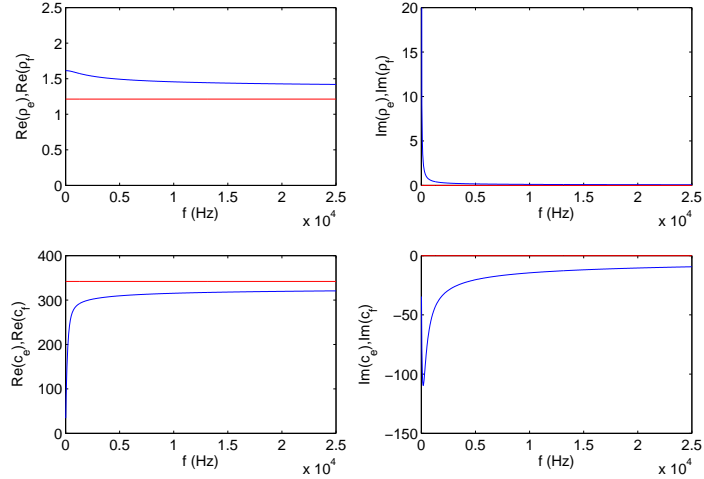}
\caption{Top left-hand panel is relative to $\Re \rho_{e}(f)$ (blue curve) and $\Re \rho_{f}(f)$ (red curve). Top right-hand panel is relative to $\Im \rho_{e}(f)$ (blue curve) and $\Im \rho_{f}(f)$ (red curve). Bottom left-hand panel is relative to $\Re c_{e}(f)$ (blue curve) and $\Re c_{f}(f)$ (red curve). Bottom right-hand panel is relative to $\Im c_{e}(f)$ (blue curve) and $\Im c_{f}(f)$ (red curve).}
\label{glv10params}
\end{center}
\end{figure}

The evolution of the JCA effective constitutive parameters with frequency $f$ is given in fig. \ref{glv10params}.
\subsection{The boundary-value problem for the acoustic response of the macroscopically-inhomogeneous layer}\label{bvp}
The total compressional wavefield in  $\Omega_{l}$ is  the scalar function  $\mathbf{u}^{[l]}(\mathbf{x},\omega)$ wherein $\mathbf{x}=(x,0,z)$. The total wavefield in the rigid medium filling $\Omega_{2}$ is nil by definition. The incident wavefield is
\begin{equation}\label{1-000}
u^{i}(\mathbf{x},\omega)=u^{[0]+}(\mathbf{x},\omega)=a^{[0]+}(\omega)\exp[i(k_{x}^{i}x+k_{z}^{i}z)]~,
\end{equation}
wherein $a^{[0]+}(\omega)$ is the spectral amplitude of the solicitation.

The plane wave nature of the solicitation and the $d$-periodicity of $\mathcal{B}$  entails the quasi-periodicity of the field, whose expression is the Floquet condition
\begin{equation}\label{1-005}
u(x+d,z,\omega)=u(x,z,\omega)\exp(ik_{x}^{i}d)~;~\forall\mathbf{x}\in \Omega_{0}+\Omega_{1}~.
\end{equation}
Consequently, as concerns the response in $\Omega_{1}$, it suffices to examine the field in $\Omega_{10}$.

The boundary-value problem in the space-frequency domain translates to the following relations (in which the superscripts $+$ and $-$ refer to the upgoing and downgoing  waves respectively) satisfied by the total displacement field $u^{[l]}(\mathbf{x};\omega)$ in $\Omega_{l}$:
\begin{equation}\label{1-010}
u^{[l]}(\mathbf{x},\omega)=u^{[l]+}(\mathbf{x},\omega)+u^{[l]-}(\mathbf{x},\omega)~;~l=0,1~,
\end{equation}
\begin{equation}\label{1-020}
u_{,xx}^{[l]}(\mathbf{x},\omega)+u_{,zz}^{[l]}(\mathbf{x},\omega)+k^{2}u^{[l]}(\mathbf{x},\omega)=0~;~\mathbf{x}\in \Omega_{l}~;~l=0,1~.
\end{equation}
\begin{equation}\label{1-030}
\frac{1}{\rho^{[1]}} u_{,z}^{[1]}(x,h,\omega)=0~;~\forall x\in [-w/2,w/2]~,
\end{equation}
\begin{equation}\label{1-033}
\frac{1}{\rho^{[0]}} u_{,z}^{[0]}(x,0,\omega)=0~;~\forall x\in [-d/2,w/2]\cup [w/2,d/2]~,
\end{equation}
\begin{equation}\label{1-035}
\frac{1}{\rho^{[1]}} u_{,x}^{[1]}(\pm w/2,z,\omega)=0~;~\forall z\in [0,h]~,
\end{equation}
\begin{equation}\label{1-040}
u^{[0]}(x,0,\omega)-u^{[1]}(x,0,\omega)=0~;~\forall x \in [-w/2,w/2]~,
\end{equation}
\begin{equation}\label{1-050}
\frac{1}{\rho^{[0]}} u_{,z}^{[0]}(x,0,\omega)-\frac{1}{\rho^{[1]}}u_{,z}^{[1]}(x,0,\omega)=0~;~\forall x \in [-w/2,w/2]~,
\end{equation}
wherein   $u_{,\zeta}$ ($u_{,\zeta\zeta}$) denotes the first (second) partial derivative of $u$ with respect to $\zeta$. Eq. (\ref{1-020}) is the space-frequency  wave equation for pressure, (\ref{1-030})-(\ref{1-035}) the expression of vanishing velocity potential  at a boundary facing a rigid medium,  (\ref{1-040}) the expression of continuity of pressure across the junction  between the $\Omega_{0}$ and the central block,  and (\ref{1-050}) the expression of continuity of velocity potential across this junction.

Since $\Omega_{0}$ is of half-infinite extent, the pressure field therein must obey the radiation condition
\begin{equation}\label{1-060}
u^{[0]-}(\mathbf{x},\omega)\sim \text{outgoing waves}~;~\|\mathbf{x}\|\rightarrow\infty~.
\end{equation}
%
\subsection{Field representations via domain decomposition and separation of variables (DD-SOV)}\label{sov}
As the preceding descriptions emphasize, it is natural to decompose $\mathbb{R}^{2}$ into three domains: $\Omega_{l}~;~l=0,1,2$, with the understanding that in $\Omega_{1}$ it is only necessary to consider what happens in the central groove subdomain.

Applying the Separation-of-Variables (SOV) technique, The Floquet condition, and the radiation condition gives rise, in the  lower domain, to the field representation:
\begin{equation}\label{2-010}
u^{[0]\pm}(\mathbf{x},\omega)=\sum_{n\in\mathbb{Z}}a_{n}^{[0]\pm}(\omega)\exp[i(k_{xn}^{[0]}x\pm k_{zn}^{[0]}z)]~,
\end{equation}
wherein:
\begin{equation}\label{2-012}
k_{xn}^{[0]}=k_{x}^{i}+\frac{2n\pi}{d}~,
\end{equation}
\begin{equation}\label{2-020}
k_{zn}^{[0]}=\sqrt{k^{2}-\left(k_{xn}^{[0]}\right)^{2}}~~;~~\Re k_{zn}^{[0]}\ge 0~~,~~\Im k_{zn}^{[0]}\ge 0~~\omega>0~,
\end{equation}
and, on account of (\ref{1-000}),
\begin{equation}\label{2-030}
a_{n}^{[0]+}(\omega)=a^{[0]+}(\omega)~\delta_{n0}~,
\end{equation}
with $\delta_{n0}$ the Kronecker delta symbol.

In the central groove,  SOV, together with the rigid boundary conditions (\ref{1-030}), (\ref{1-035}), lead to
\begin{equation}\label{2-040}
u^{[1]\pm}(\mathbf{x},\omega)=\sum_{m=0}^{\infty}a_{m}^{[1]}(\omega)\cos[k_{xm}^{[1]}(x+w/2)]\exp[\pm k_{zm}^{[1]}(z-h)]~,
\end{equation}
in which
\begin{equation}\label{2-050}
k_{xm}^{[1]}=\frac{m\pi}{w}~,
\end{equation}
\begin{equation}\label{2-060}
k_{zm}^{[1}=\sqrt{k^{2}-\big(k_{xm}^{[1]}\big)^{2}}~~;~~\Re k_{zm}^{[1]}\ge 0~~,~~\Im k_{zm}^{[1]}\ge 0~~\omega>0~.
\end{equation}
%
\subsection{Exact solutions for the unknown coefficients}
Eqs. (\ref{1-033}) and (\ref{1-050})  entail
\begin{equation}\label{3-003}
\frac{1}{\rho^{[0]}}\int_{-d/2}^{d/2}u^{[0]}_{,z}(x,0,\omega)\exp(-ik_{xj}^{[0]}x)\frac{dx}{d}=
\frac{1}{\rho^{[1]}}\int_{-w/2}^{w/2}u^{[01}_{,z}(x,0,\omega)\exp(-ik_{xj}^{[0]}x)\frac{dx}{d}~;~\forall j=0,\pm 1,\pm 2,....~,
\end{equation}
which, on account of the SOV field representations and the identity
\begin{equation}\label{3-005}
\int_{-d/2}^{d/2}\exp\left[i\left(k_{xn}^{[0]}-k_{xj}^{[0]}\right)x\right]\frac{dx}{d}=\delta_{nj}~,
\end{equation}
($\delta_{nj}$ is the Kronecker delta) yields
\begin{equation}\label{3-010}
a_{j}^{[0]-}=a_{j}^{[0]+}-\frac{w}{2id}\frac{\rho^{[0]}}{\rho^{[1]}}\frac{1}{k_{zj}^{[0]}}
\sum_{m=0}^{\infty}a_{m}^{[1]}k_{zm}^{[1]}\sin\left(k_{zm}^{[1]}h\right)E_{jm}^{-}~;~\forall j=0,\pm 1,\pm 2,....
~,
\end{equation}
wherein
\begin{multline}\label{3-040}
E_{jm}^{\pm}=\int_{-w/2}^{w/2}\exp\left(\pm ik_{xj}^{[0]}x\right)\cos\left[k_{xm}^{[1]}(x+w/2)\right]\frac{dx}{w/2}=\\
i^{m}\left\{\text{sinc}\left[\left(\pm k_{xj}^{[0]}+k_{xm}^{[1]}\right)w/2\right]+(-1)^{m}\text{sinc}\left[\left(\pm k_{xj}^{[0]}-k_{xm}^{[1]}\right)w/2\right]\right\}
~,
\end{multline}
with sinc$(\zeta)=\frac{\sin\zeta}{\zeta}$ and sinc(0)=1.

Eq. (\ref{1-040})  entails
\begin{multline}\label{3-050}
\int_{-w/2}^{w/2}u^{[[0]}(x,0,\omega)\cos\left[k_{xl}({[1]}(x+w/2)\right]\frac{dx}{w/2}=\\
\int_{-w/2}^{w/2}u^{[[1]}(x,0,\omega)\cos\left[k_{xl}({[1]}(x+w/2)\right]\frac{dx}{w/2}~;~\forall l=0,1,2,....~,
\end{multline}
which, on account of the SOV field representations, and the identity
\begin{equation}\label{3-055}
\int_{-w/2}^{w/2}\cos\left[k_{xm}^{[1]}(x+w/2)\right]\cos\left[k_{xl}^{[1]}(x+w/2)\right]=\frac{2}{\epsilon_{l}}\delta_{lm}~,
\end{equation}
with $\epsilon_{l}$ the Neumann symbol (=1 for $l=0$ and =2 for $l>0$), enables us to find
\begin{equation}\label{3-060}
a_{l}^{[1]}=\left(\frac{\epsilon_{l}}{2\cos\left(k_{zl}^{[1]}h\right)}\right)
\sum_{n-\infty}^{\infty}\left(a_{n}^{[0]+}+a_{n}^{[0]-}\right)E_{nl}^{+}~;~\forall l=0,1,2,....
\end{equation}
We thus have at our disposal two coupled expressions (i.e., (\ref{3-010}) and (\ref{3-060}) which make it possible to determine the two sets of unknowns $\{a_{n}^{[0]-}\}$, $\{a_{n}^{[1]}\}$. Note that the number of members of each of these sets is infinite which is the fundamental source of  complexity of the problem at hand and the principal reason why one should strive to simplify the theoretical analysis. This will be done in a later section.
\subsection{Linear system for the  set of unknown coefficients}
Inserting (\ref{3-010}) into(\ref{3-060})  yields, after the summation interchange, the system of linear equations:
\begin{equation}\label{4-040}
\sum_{m=0}^{\infty}X_{lm}Y_{m}=Z_{l}~;~\forall l=0,1,2,....~,
\end{equation}
wherein
\begin{equation}\label{4-045}
Y_{m}=a_{m}^{[1]}~,~~Z_{l}=a^{[0]+}\epsilon_{l}E_{0l}^{+}~,
\end{equation}
\begin{equation}\label{4-050}
X_{lm}=\delta_{lm}\cos\left(k_{zm}^{[1]}h\right)+\frac{w}{2id}\frac{\rho^{[0]}}{\rho^{[1]}}\frac{\epsilon_{l}}{2}k_{zm}^{[1]}\sin\left(k_{zm}^{[1]}h\right)\Sigma_{lm}~~,~~
\Sigma_{lm}=\sum_{n=-\infty}^{\infty}\frac{1}{k_{zn}^{[0]}}E_{nl}^{+}E_{nm}^{-}~.
\end{equation}
Once the $Y_{m}=a_{m}^{[1]}$ are determined they can be inserted into (\ref{3-010}) to determine the $a_{j}^{[0]-}$, i.e.,
\begin{equation}\label{4-055}
a_{j}^{[0]-}=a_{j}^{[0]+}-\frac{w}{2id}\frac{\rho^{[0]}}{\rho^{[1]}}\frac{1}{k_{zj}^{[0]}}
\sum_{m=0}^{\infty}Y_{m}k_{zm}^{[1]}\sin\left(k_{zm}^{[1]}h\right)E_{jm}^{-}~;~\forall j=0,\pm 1,\pm 2,....
~,
\end{equation}

Until now everything has been rigorous provided the equations in the statement of the boundary-value problem are accepted as the true expression of what is involved in the  acoustic wave response of our grating and certain summation interchanges are valid. In order to actually solve for the sets $\{a_{n}^{[0]-}\}$ and $\{a_{m}^{[1]}\}$ (each of whose populations is  considered to be infinite at this stage) we must resort either to numerics or to approximations.
\subsection{Numerical issues concerning the  system of equations for $\{a_{m}^{[1]}\}$}
We strive to obtain numerically the set $\{a_{m}^{[1]}\}$ from the linear system of equations (\ref{4-040}). Once this set is found, it is introduced into (\ref{3-010}) to obtain the set $\{a_{n}^{[0]-}\}$.  When all these coefficients (we mean those whose values depart significantly from zero) are found, they enable the computation of the acoustic wave response (i.e., the displacement field) in all the subdomains of the configuration via (\ref{1-000}), (\ref{1-010}), (\ref{2-010}), (\ref{2-040}).

Concerning the resolution of the infinite system of linear equations (\ref{4-040}), the procedure is basically to replace it by the finite system of linear equations
\begin{equation}\label{5-010}
\sum_{m=0}^{M}X^{(M)}_{lm}Y_{m}^{(M)}=Z_{l}~;~l=0,1,2,...M~,
\end{equation}
in which $X^{(M)}_{lm}$ signifies that the series in $X_{lm}$ is limited to the terms $n=0,\pm 1,...,\pm M$,  and to increase $M$ so as to generate the sequence of numerical solutions $\{Y_{m}^{(0)}\}$, $\{Y_{m}^{(1)},Y_{m}^{(2)}\}$,....until the values of the first few members of  of these sets stabilize and the remaining members become very small (this is the so-called 'reduction method' \cite{ri13} of resolution of an infinite system of linear equations).

Note that to each $Y_{m}^{(M)}=a_{m}^{[1](M)}$ is associated $a_{m}^{[0]-(M)}$ via (\ref{4-055}), i.e.,
\begin{equation}\label{5-015}
a_{j}^{[0]-(M)}=a_{j}^{[0]+}-\frac{w}{2id}\frac{\rho^{[0]}}{\rho^{[1]}}\frac{1}{k_{zj}^{[0]}}
\sum_{m=0}^{M}Y_{m}^{(M)}k_{zm}^{[1]}\sin\left(k_{zm}^{[1]}h\right)E_{jm}^{-}~;~\forall j=0,\pm 1,\pm 2,....\pm M
~,
\end{equation}

The so-obtained numerical solutions (it being implicit that $Y_{m}^{(M)}=a^{[1](M)}=0~;~m>M$ and $a_{j}^{[0]-(M)}~;~|j|>M$), which for all practical purposes can be considered as 'exact' for sufficiently-large $M$ (of the order of 25 for the range of frequencies and grating  parameters considered herein) and which are in agreement with numerical results obtained by a finite element method \cite{gr05}, constitute the reference by which we shall measure the accuracy of the approximate solutions of later sections.
\subsection{Conservation of flux for the periodic structure}\label{fluxper}
We again refer to fig. \ref{blockabsgrating} wherein we now focus on the integration domain $\Omega_{00}\subset\Omega_{0}$   bounded by the dashed lines in the lower part of the figure.

Since the boundary-value problem is the same as previously, we again refer to  its governing equations given in sect. \ref{bvp}. Let $\boldsymbol{\nu}$ designate the outward-pointing unit normal to a domain $\Omega$ whose closed boundary is $\partial\Omega$, and assume that the total pressure field within $\Omega$ is $u(\mathbf{x},\omega)$ obeying the Helmholtz equation $(\Delta+k^{2})u=0$. Then, applying Green's second identity leads to
\begin{equation}\label{6-010}
\Im\int_{\partial\Omega}u^{*}\boldsymbol{\nu}\cdot\nabla u ~d\gamma+\Im[k^{2}]\int_{\Omega}\|u\|^{2}d\varpi=0
~,
\end{equation}
wherein $*$ designates the complex conjugate operator, $d\gamma$ the differential element of arc length, and $d\varpi$ the differential element of area.
Since, by definition, it was assumed that air is lossless, $\Im k^{[0]}=0$, so that
\begin{equation}\label{6-020}
\Im\int_{\partial\Omega_{00}}u^{[0]*}\boldsymbol{\nu}\cdot\nabla u^{[0]} ~d\gamma=0
~,
\end{equation}
and, since the field obeys the Floquet condition,
\begin{equation}\label{6-030}
-\Im\int_{-d/2}^{d/2}u^{[0]*}(x,-H,\omega)u^{[0]}_{,z}(x,-H,\omega)~dx+\Im\int_{-d/2}^{d/2}u^{[0]*}(x,0,\omega)u^{[0]}_{,z}(x,0,\omega)~dx=0
~.
\end{equation}
Due to the boundary and transmission conditions on  $z=0$, we obtain
\begin{equation}\label{6-040}
-\Im\int_{-d/2}^{d/2}u^{[0]*}(x,-H,\omega)u^{[0]}_{,z}(x,-H,\omega)~dx+
\Im\left[\frac{\rho^{[0]}}{\rho^{[1]}}\int_{-w/2}^{w/2}u^{[1]*}(x,0,\omega)u^{[1]}_{,z}(x,0,\omega)~dx\right]=0
~.
\end{equation}
By employment of the plane-wave field representations in $\Omega_{00}$ we find that the first term does not depend on $H$, and, in fact:
\begin{equation}\label{6-050}
-k_{z0}^{[0]}d\|a^{[0]+}\|^{2}-\Re\sum_{n\in\mathbb{Z}}k_{zn}^{[0]}d\|a^{[0]-}\|^{2}+
\Im\left[\frac{\rho^{[0]}}{\rho^{[1]}}\int_{-w/2}^{w/2}u^{[1]*}(x,0,\omega)u^{[1]}_{,z}(x,0,\omega)~dx\right]=0
~,
\end{equation}
from which it follows that
\begin{equation}\label{6-060}
\rho(\omega)+\alpha(\omega)=1
~.
\end{equation}
in which:
\begin{equation}\label{6-070}
\rho(\omega)=\Re\sum_{n\in\mathbb{Z}}\frac{k_{zn}^{[0]}}{k_{z0}^{[0]}}\left\|\frac{a_{n}^{[0]-}}{a^{[0]+}}\right\|^{2}
~,
\end{equation}
\begin{equation}\label{6-080}
\alpha(\omega)=
\Im\left[\left(\frac{\rho^{[0]}}{\rho^{[1]}}\right)
\frac{1}{\|a^{[0]+}\|^{2}}
\int_{-w/2}^{w/2}u^{[1]*}(x,0,\omega)u^{[1]}_{,z}(x,0,\omega)~\frac{dx}{k_{z0}^{[0]}d}\right]
~,
\end{equation}
By introducing the SOV field representation relative to $u^{[1]}$ into (\ref{6-080}) we finally get
\begin{equation}\label{6-090}
\alpha(\omega)=
\Im\left[\frac{\rho^{[0]}}{\rho^{[1]}}\frac{w}{d}
\sum_{m=0}^{\infty}
\frac{k_{zm}^{[1]}}{k_{z0}^{[0]}}\left\|\frac{a_{m}^{[1]}}{a^{[0]+}}\right\|^{2}
\left(\cos(k_{zm}^{[1]}h)\right)^{*}\sin(k_{zm}^{[1]}h)\right]~.
\end{equation}

Eq. (\ref{6-070}) shows us that $\rho(\omega)$ depends only on the 'reflected' field  in the half space $\Omega_{0}$  so that it is legitimate to associate it with what in   optics \cite{ly77} is termed the  'spectral reflectance'. We prefer to term it the (normalized) 'reflected flux'. Eq. (\ref{6-090}) shows us that $\alpha(\omega)$ depends only on the field in the foam-filled grooves (the foam playing the role of absorbing medium),  so that it is legitimate to associate it with what in optics is termed the   'spectral absorptance'. We prefer to call it the (normalized)  'absorbed flux'. It follows that (\ref{6-060}) is the expression of the conservation of (normalized) flux, the left-hand side of this equation representing the (normalized) output flux and the right hand side the (normalized) input flux.

The acoustic design problem, if such be our preoccupation, is to maximize, via the presence of the grating, $\alpha(\omega)$ over the widest possible low-frequency bandwidth; (\ref{6-060}) shows us that this is possible only by reducing  $\rho$ as close as possible to zero at which point $\alpha$ attains its maximal value of unity. More specifically, we shall show that the absorbed flux of the foam-loaded rigid grating structure can be larger, over a certain low-frequency bandwidth) than the absorbed flux of a  layer entirely filled with foam and backed by a rigid material when the foams and thicknesses of both structures are identical.

As we shall employ further on the $M<\infty$ approximations of the field amplitudes, we must also define the finite-$M$ approximations of the reflected flux ($\rho^{(M)}$) and absorbed flux ($\alpha^{(M)}$). Quite naturally, these are:
\begin{equation}\label{6-092}
\rho^{(M)}(\omega)=\Re\sum_{n=-M}^{M}\frac{k_{zn}^{[0]}}{k_{z0}^{[0]}}\left\|\frac{a_{n}^{[0]-(M)}}{a^{[0]+}}\right\|^{2}
~,
\end{equation}
\begin{equation}\label{6-094}
\alpha^{(M)}(\omega)=
\Im\left[\frac{\rho^{[0]}}{\rho^{[1]}}\frac{w}{d}
\sum_{m=0}^{M}
\frac{k_{zm}^{[1]}}{k_{z0}^{[0]}}\left\|\frac{a_{m}^{[1](M)}}{a^{[0]+}}\right\|^{2}
\left(\cos(k_{zm}^{[1]}h)\right)^{*}\sin(k_{zm}^{[1]}h)\right]~.
\end{equation}
These approximations (if they are satisfactory) should satisfy the conservation relation
\begin{equation}\label{6-096}
\rho^{(M)}(\omega)+\alpha^{(M)}(\omega)=1~.
\end{equation}
%
\subsection{Numerical comparison of the $M=0,1,2$ approximations of the grating structure response to their exact counterpart for the filler material in \cite{gdd11}}\label{M012}
We are now in a position to compare the noteworthy features (amplitudes of the fields in the air and in the porous filler, reflected flux, absorbed flux and output flux) of the $M=0,1,2$ approximations of the grating response to the exact (actually the $M=25$ numerical approximation of this) response.

\begin{figure}[ht]
\begin{center}
\includegraphics[width=0.7\textwidth]{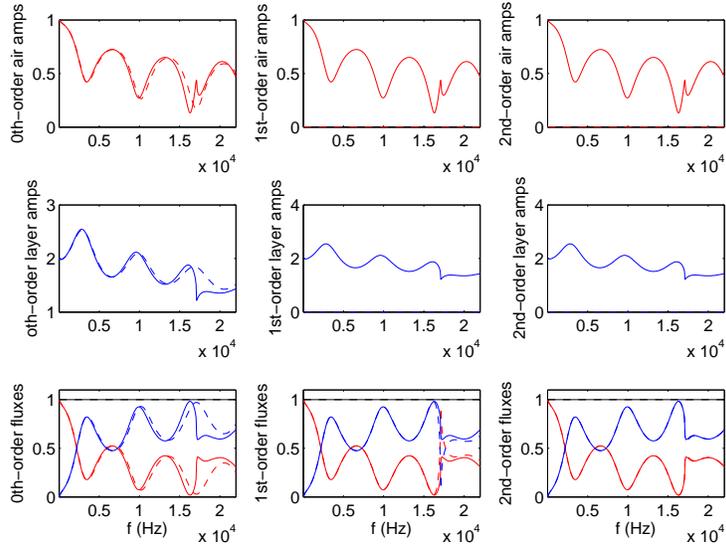}
\caption{The reference solutions (full blue or red curves) are for $M=25$. The dashed (blue or red) curves in the left-hand, middle and right-hand panels are for $M=0,1,2$ respectively. The upper row of panels depict the moduli of the amplitudes ($\|a_{0}^{[0]-(M)}\|$) in the air-filled half space, the middle row of panels are for the moduli of the pressure field amplitudes in the layer region ($\|a_{0}^{[1](M)}\|$) and the lower row of panels are for the reflected (red curves, $\rho^{(M)}$), absorbed (blue curves, $\alpha^{(M)}$) and output  (black curves, $\rho^{(M)}+\alpha^{(M)}$) fluxes.  Case  $h=0.02~m$,  $w=0.015~m$, $d=0.02~m$,  $\theta^{i}=0^{\circ}$.}
\label{m012-05}
\end{center}
\end{figure}

In figs. \ref{m012-05}-\ref{m012-08}, the  reference solutions (full blue or red curves) are for $M=25$ which are compared to the approximate  (dashed curves) $M=0$ solutions (left-hand  panels), $M=1$ solutions (middle panels) and $M=2$ solutions (right-hand panels). The upper row of panels are for the moduli of the pressure field amplitudes in the air half space, the middle row of panels are for the moduli of the pressure field amplitudes in the layer region and the lower row of panels are for the reflected (red curves), absorbed (blue curves) and output (black curves) fluxes. Note that the output flux is taken as the sum of the computed reflected flux and computed absorbed flux and should be equal to 1 in order for flux to be conserved.

\begin{figure}[ptb]
\begin{center}
\includegraphics[width=0.7\textwidth]{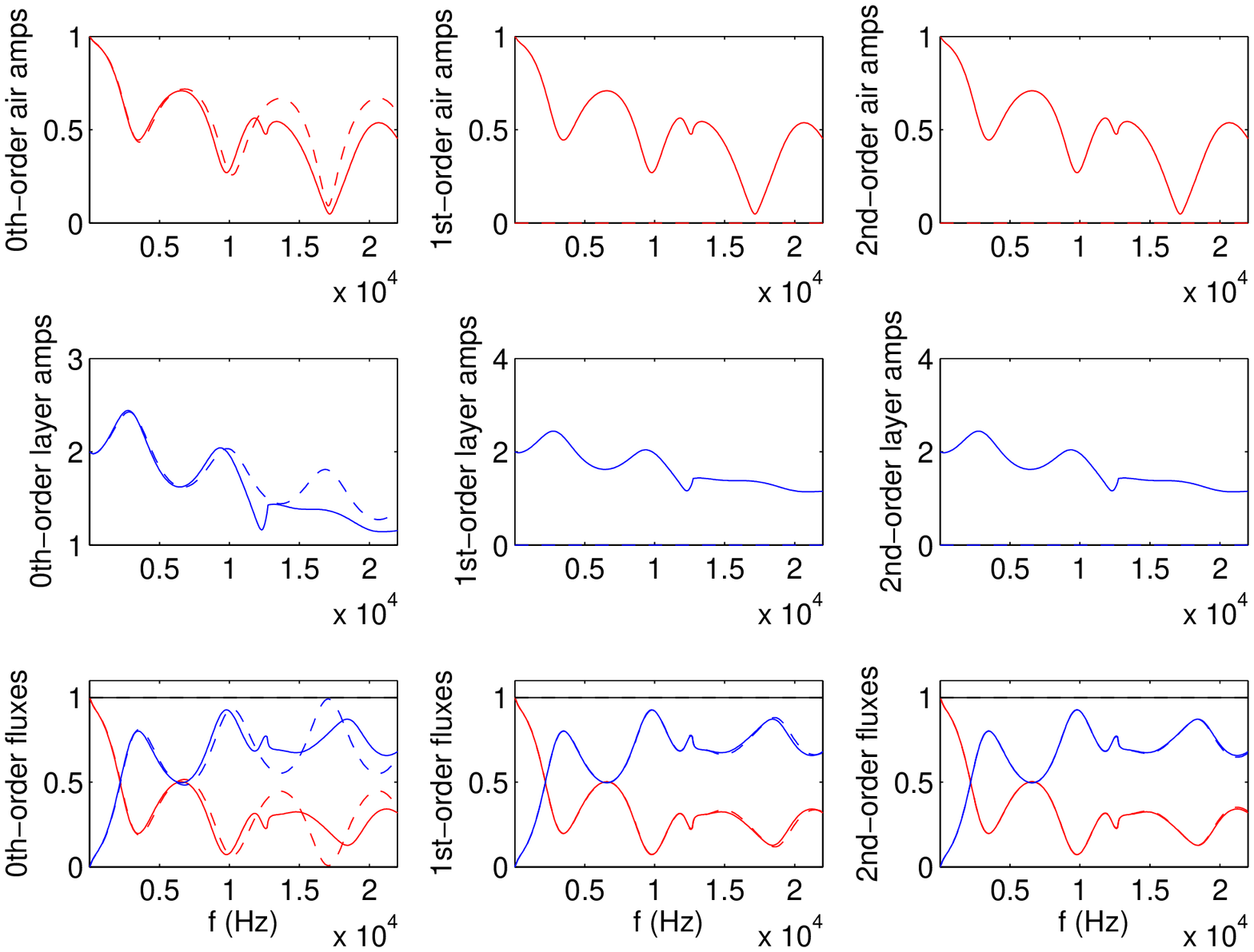}
\caption{Same as fig. \ref{m012-05} except that $\theta^{i}=20^{\circ}$.}
\label{m012-06}
\end{center}
\end{figure}
\begin{figure}[ptb]
\begin{center}
\includegraphics[width=0.7\textwidth]{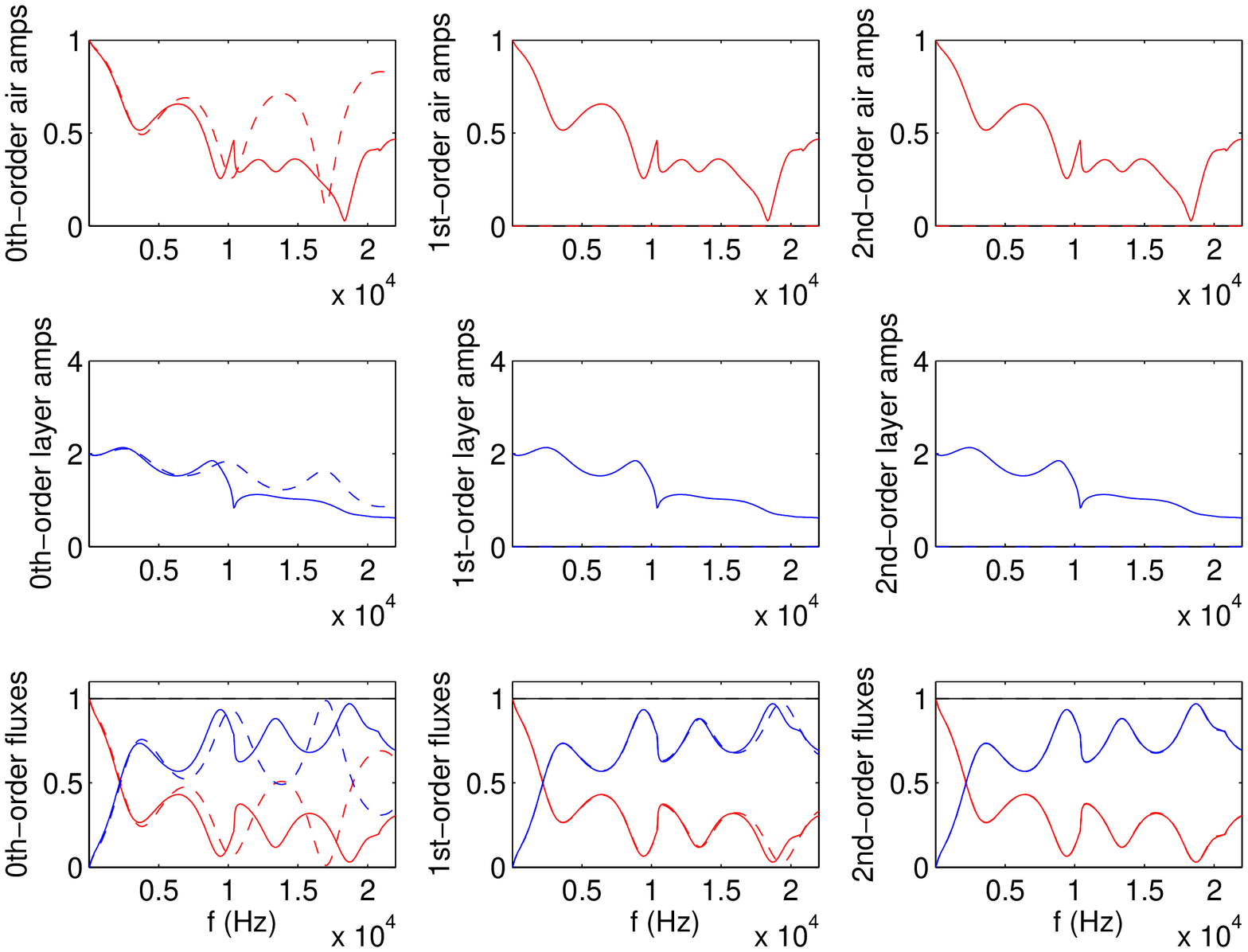}
\caption{Same as fig. \ref{m012-05} except that $\theta^{i}=40^{\circ}$.}
\label{m012-06}
\label{m012-07}
\end{center}
\end{figure}
\begin{figure}[ptb]
\begin{center}
\includegraphics[width=0.7\textwidth]{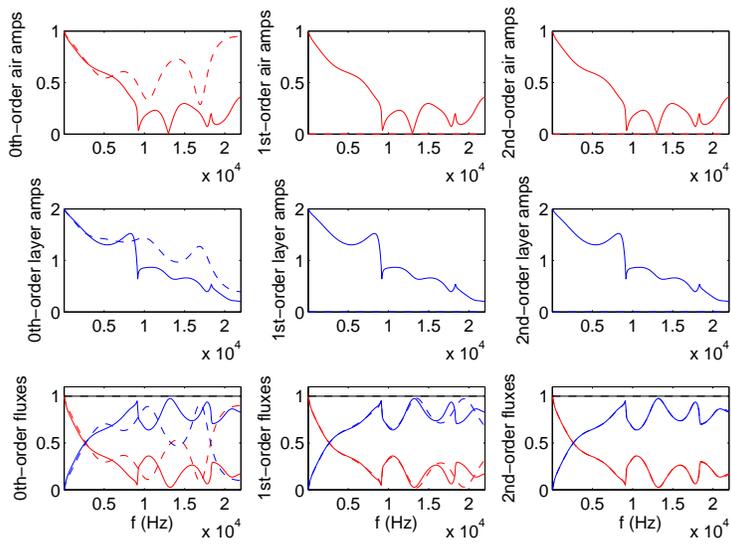}
\caption{Same as fig. \ref{m012-05} except that $\theta^{i}=60^{\circ}$.}
\label{m012-06}
\label{m012-08}
\end{center}
\end{figure}
\clearpage
\newpage

The figures show that the $M=0$ solution is a good approximation of the response functions at low frequencies provided the absolute value of the incident angle does not exceed $40^{\circ}$. Moreover, all the (i.e., $M=0,1,2$) approximate solutions, as well as the $M=25$ reference solutions satisfy (numerically) the conservation of flux relation. Consequently, it seems reasonable to adopt the $M=0$ solution as a suitable representation of the low-frequency, near normal-incidence, response of the grating. This is what is done in the sequel.
\section{Response of a macroscopically-homogeneous lossy layer to airborne sound}\label{fluxlay}\label{cflayr}
%
\subsection{Why study this problem?}\label{fluxlay}\label{cflayr}
The reason why we should be interested in this problem is because the configuration of a rigidly-backed layer entirely filled with foam is ubiquitous in the noise-reduction applications and thus serves as a reference to test the effectiveness of replacing the layer by a grating of alternating rigid and foam materials. Also, further on, we shall shown that the low-frequency, small incident angle, response of the inhomogeneous layer reduces to that of avhomogeneous layer.
\subsection{Description of the configuration}\label{fluxlay}\label{cflayr}
The first task is to establish the quantities that provide a measure of the noise-reduction effectiveness of the introduction of the foam layer between the air and rigid half spaces. We shall also show that these quantities are related by a conservation law.
\begin{figure}[ht]
\begin{center}
\includegraphics[width=0.65\textwidth]{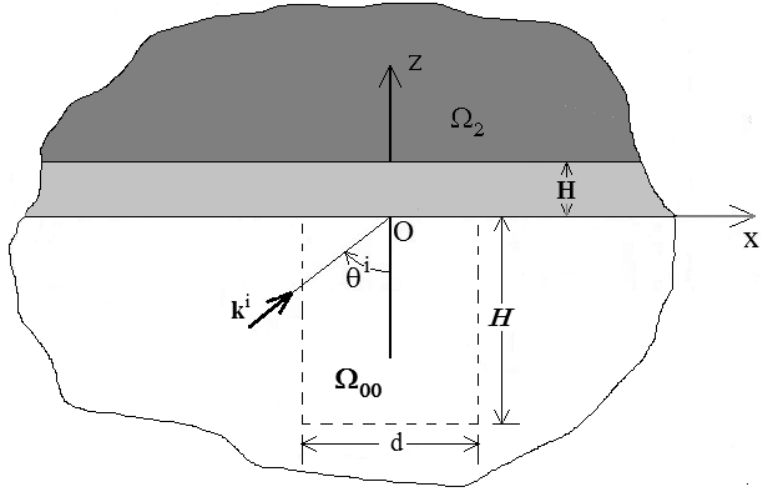}
\caption{Same as fig. \ref{blockabsgrating} except that the former macroscopically periodically-heterogeneous layer of thickness $h$ (i.e., the grating whose grooves are filled with foam of density $\rho^{[1]}$ and wavespeed $c^{[1]}$) is now a macroscopically-homogenous layer of thickness H filled with another foam of density $R^{[1]}$ and wavespeed $C^{[1]}$. Moverover, the lower integration domain $\Omega_{00}$ is as previously.}
\label{blockabslayrflux}
\end{center}
\end{figure}

Fig. \ref{blockabslayrflux} depicts the problem  in the sagittal plane in which: $\Omega_{0}$ is the half-space domain occupied by a light fluid such as air,   $\Omega_{1}$  the domain of the layer occupied by another macroscopically-homogeneous fluid (i.e., the approximation resulting from the JCA model of foam) which is  lossy and dispersive,  and $\Omega_{2}$ the  half-space above the layer occupied by an acoustically-rigid material.

The densities of the lower  medium and layer are $R^{[0]}$ and $R^{[1]}$ respectively, with  $R^{[0]}$  positive real and $R^{[1]}$ complex (real and imaginary parts positive).  The compressional-wave velocities in the lower medium and layer are $C^{[0]}$ and $C^{[1]}$ respectively, with  $C^{[0]}$  positive real and $C^{[1]}$ complex (real part positive, imaginary part negative).

The  light fluid-borne plane-wave solicitation is as previously and given by
\begin{equation}\label{7-020}
U^{i}(\mathbf{x},\omega)=U^{[0]+}(\mathbf{x},\omega)=A^{[0]+}(\omega)\exp[i(K_{x}^{i}x+K_{z}^{i}z)]~,
\end{equation}
wherein $A^{[0]+}(\omega)$ is the spectral amplitude of the solicitation, $K_{x}^{i}=K^{[0]}\sin\theta^{i}$, $K^{i}_{z}=K^{[0]}\cos\theta^{i}$, and $K^{[l]}=\omega/C^{[l]}$.  The total wavefield $U(\mathbf{x},\omega)$ in $\Omega_{l}$ is designated by $U^{[l]}(\mathbf{x},\omega)$.
\subsection{The boundary-value problem of the response of the rigidly-backed layer structure to a plane wave}\label{bvplayr}
The boundary-value problem in the space-frequency domain translates to the following relations:
\begin{equation}\label{7-030}
U^{[l]}(\mathbf{x},\omega)=U^{[l]+}(\mathbf{x},\omega)+U^{[l]-}(\mathbf{x},\omega)~;~l=0,1~,
\end{equation}
\begin{equation}\label{7-040}
U_{,xx}^{[l]}(\mathbf{x},\omega)+U_{,zz}^{[l]}(\mathbf{x},\omega)+(K^{[l]})^{2}U^{[l]}(\mathbf{x},\omega)=0~;~\mathbf{x}\in \Omega_{l}~;~l=0,1~.
\end{equation}
\begin{equation}\label{7-050}
\frac{1}{R^{[1]}}U_{,z}^{[1]}(x,{\text H},\omega)=0~;~\forall x\in \mathbb{R}~,
\end{equation}
\begin{equation}\label{7-060}
U^{[0]}(x,0,\omega)-U^{[1]}(x,0,\omega)=0~;~\forall x \in \mathbb{R},
\end{equation}
\begin{equation}\label{7-070}
\frac{1}{R^{[0]}}(x,0,\omega)-\frac{1}{R^{[1]}}U_{,z}^{[1]}(x,0,\omega)=0~;~\forall x \in \mathbb{R}~.
\end{equation}
\begin{equation}\label{7-080}
U^{[0]-}(\mathbf{x},\omega)\sim \text{outgoing waves}~;~\mathbf{x}\rightarrow\infty~.
\end{equation}
Due to the translational symmetry along $x$, the field obeys a sort of Floquet condition
\begin{equation}\label{7-082}
U^{[l]}(x+d,z,\omega)=U^{[l]}(x,z,\omega)\exp(iK_{x}^{i}d)~;~l=0,1~,
\end{equation}
this relation being true for all $d$.
\subsection{Conservation of flux}
We proceed as in sect. \ref{fluxper} to obtain, by means of the boundary condition, the transmission conditions, and the 'Floquet' condition
\begin{equation}\label{7-084}
-\Im\int_{-d/2}^{d/2}U^{[0]*}(x,-H,\omega)U{[0]}_{,z}(x,-H,\omega)~dx+
\Im\left[\frac{R^{[0]}}{R^{[1]}}\int_{-d/2}^{d/2}U^{[1]*}(x,0,\omega)U^{[1]}_{,z}(x,0,\omega)~dx\right]=0
~.
\end{equation}
To make this relation more explicit, we apply the DD-SOV technique, and the radiation condition to obtain, in the  lower domain, the field representation:
\begin{equation}\label{7-090}
U^{[0]\pm}(\mathbf{x},\omega)=A^{[0]\pm}(\omega)\exp[i(K_{x}^{[0]}x\pm K_{z}^{[0]}z)]~,
\end{equation}
wherein:
\begin{equation}\label{7-100}
K_{x}^{[0]}=K_{x}^{i}~,
\end{equation}
\begin{equation}\label{7-120}
K_{z}^{[0]}=\sqrt{\big(K^{[0]}\big)^{2}-\left(K_{x}^{[0]}\right)^{2}}~~;~~\Re K_{z}^{[0]}\ge 0~~,~~\Im K_{z}^{[0]}\ge 0~~\omega>0~.
\end{equation}

In the layer, the SOV, together with the  boundary condition (\ref{7-050}),  lead to
\begin{equation}\label{7-140}
U^{[1]}(\mathbf{x},\omega)=A^{[1]}(\omega)\exp\left[iK_{x}^{[1]}x\right]\cos\left[K_{z}^{[1]}(z-{\text H})\right]~,
\end{equation}
in which
\begin{equation}\label{7-150}
K_{x}^{[1]}=K_{x}^{[0]}=K_{x}^{i}~,
\end{equation}
\begin{equation}\label{7-160}
K_{z}^{[1]}=\sqrt{\big(K^{[1]}\big)^{2}-\big(K_{x}^{[1]}\big)^{2}}~~;~~\Re K_{z}^{[1]}\ge 0~~,~~\Im K_{z}^{[1]}\ge 0~~\omega>0~.
\end{equation}
Employing these relations in (\ref{7-084}) leads, in the same manner as previously, to the conservation of flux relation
\begin{equation}\label{7-162}
\mathcal{R}(\omega)+\mathcal{A}(\omega)=1~,
\end{equation}
in which
\begin{equation}\label{7-164}
\mathcal{R}(\omega)=\left\|\frac{A^{[0]-}}{A^{[0]+}}\right\|^{2}
~,
\end{equation}
\begin{equation}\label{7-166}
\mathcal{A}(\omega)=
\Im\left[
\left\|
\frac{A^{[1]}}{A^{[0]+}}
\right\|^{2}\left(\frac{R^{[0]}K_{z}^{[1]}}{R^{[1]}K_{z}^{[0]}}\right)
\left(\cos(K_{z}^{[1]}{\text H})\right)^{*}\sin(K_{z}^{[1]}{\text H})\right]=\left\|
\frac{A^{[1]}}{A^{[0]+}}
\right\|^{2}E(\omega)
~.
\end{equation}
These expressions show, as one would expect, that the reflected flux $\mathcal{R}$ and absorbed flux $\mathcal{A}$ do not depend  on  $d$ (the width of the integration domain) nor on $H$ (the height of the  integration domain). Furthermore, although $\mathcal{R}$ is a straightforward function of  the reflected  field amplitude via $\|A^{[0]-}\|^{2}$, $\mathcal{A}$ is not a straightforward function of the amplitude of the field in the layer via $\|A^{[1]}\|^{2}$ because this latter function is multiplied by $E$. To go further into this matter, we must solve for these two amplitudes.
\subsection{Exact solution for the unknown coefficients}\label{layramps}
The introduction of the field representations into  (\ref{7-060})-(\ref{7-070}) yields the two equations
\begin{equation}\label{7-170}
A^{[0]+}+A^{[0]-}=A^{[1]}\cos\left(-K_{z}^{[1]}{\text H}\right)~,
\end{equation}
\begin{equation}\label{7-180}
\frac{i}{R^{[0]}}K_{z}^{[0]}\left(A^{[0]+}-A^{[0]-}\right)=-\frac{1}{R^{[1]}}K_{z}^{[1]}A^{[1]}\sin\left(-K_{z}^{[1]}{\text H}\right)~,
\end{equation}
the exact solution of which is:
\begin{equation}\label{7-190}
A^{[1]}=A^{[0]+}\left[\frac{2}
{\cos\left(K_{z}^{[1]}H\right)+\frac{R^{[0]}K_{z}^{[1]}}{iR^{[1]}K_{z}^{[0]}}\sin\left(K_{z}^{[1]}{\text H}\right)}\right]=
A^{[0]+}\left[\frac{\mathcal{N}^{[1]}}{\mathcal{D}}\right]~.
\end{equation}
\begin{equation}\label{7-195}
A^{[0]-}=A^{[0]+}\left[\frac{\cos\left(K_{z}^{[1]}{\text H}\right)-\frac{R^{[0]}K_{z}^{[1]}}{iR^{[1]}K_{z}^{[0]}}\sin\left(K_{z}^{[1]}{\text H}\right)}
{\cos\left(K_{z}^{[1]}{\text H}\right)+\frac{R^{[0]}K_{z}^{[1]}}{iR^{[1]}K_{z}^{[0]}}\sin\left(K_{z}^{[1]}{\text H}\right)}\right]=
A^{[0]+}\left[\frac{\mathcal{N}^{[0]}}
{\mathcal{D}}\right]~,
\end{equation}
wherein
\begin{equation}\label{7-197}
C=\cos\left(K_{z}^{[1]}{\text H}\right)~,~~S=\sin\left(K_{z}^{[1]}{\text H}\right)~,~~G=\frac{R^{[0]}K_{z}^{[1]}}{R^{[1]}K_{z}^{[0]}}~,~~\mathcal{D}=C-iGS~,~~\mathcal{N}^{[1]}=2~,~~\mathcal{N}^{[0]}=C+iGS
~.
\end{equation}
It follows that:
\begin{equation}\label{7-198}
\mathcal{R}(\omega)=\frac{\|C\|^{2}+\|GS\|^{2}-2\Im(GSC^{*})}{\|C\|^{2}+\|GS\|^{2}+2\Im(GSC^{*})}~, ~~\mathcal{A}(\omega)=\frac{4\Im(GSC^{*})}{\|C\|^{2}+\|GS\|^{2}+2\Im(GSC^{*})}~,
\end{equation}
which shows that (\ref{7-162}) is satisfied, i.e., flux is conserved, as is necessary.

Eq. (\ref{7-198}) also shows that there can be no absorption unless $G$ and/or $SC^{*}$ are complex which means that non-vanishing absorption requires that $R^{[1]}$ and/or $C^{[1]}$ be complex. Choosing the values of these complex parameters to obtain maximal absorption is a multiparameter optimization problem that will be considered further on. Also, we shall then give numerical examples of the way the absorbed flux varies with the thickness and constitutive parameters of the foam layer as well as with the frequency.
\subsection{Origin and meaning of the so-called quarter-wavelength resonances}\label{qwr}
The so-called quarter-wavelength resonances (QWR) are often mentioned (without giving their theoretical basis) in connection with explanations of the oscillatory behavior in general, and the  maxima in particular, of  the response of a rigidly-backed (usually homogeneous) layer submitted to airborne sound  \cite{gdd11,jrg18}.  At present, we address three questions relative to the QWR: (i) where do they come from?, (ii) are they really resonances?, and (iii) what  does the theory underlying the QWR actually tell us about the response of the layer?

In elastic wave problems, particularly those related to studies of ground shaking during earthquakes in a configuration in which the ground overlies a relatively-soft soil layer, whose boundaries are planar and mutually-parallel, and which is underlain by a very hard rock basement, the notion of soil layer (natural) frequency was developed as early as 1930 \cite{sk30} and employed successfully by scores of geophysicists to furnish simple explanations of many features (frequencies of occurrence and heights of the  peaks of the transfer functions) of recorded seismograms on the earth's surface. The notion of soil frequency was even employed (and renamed the 1D resonance frequency) to furnish a rough explanation of the seismic response of surface layers with curved upper or lower boundaries (hills or basins), e.g. \cite{al70,bb85,wi95}. It is probable that similar notions appeared even earlier than 1930 in connection with optical problems involving a dielectric layer over a near-perfect conductor and perhaps in similar water wave contexts. Here, we adopt the acoustic wave formalism for a homogeneous layer, with flat-plane parallel boundaries, backed above by an infinitely-rigid half-space and below by a light fluid such as air in which propagates a plane body acoustic wave. The Neumann boundary condition on the upper ($z=H$) face of the layer is the same as the one which prevails on the ground overlying the geophysical flat-faced layer so that we shall proceed as in the geophysical problem by focusing our attention (there on the ground) on the field on the upper face (here in contact with the rigid backing and termed the 'top') of the layer.

We showed previously that the frequency domain pressure in the layer is:
\begin{equation}\label{7-310}
U^{[1]}(x,z,f)=A^{[1]}\cos[K_{z}^{[1]}(z-H)]\exp(iK_{x}x)~,
\end{equation}
wherein $K_{x}=K_{x}^{[0]}=K_{x}^{[1]}$ and
\begin{equation}\label{7-310}
A^{[1]}=A^{[0]+}\left(\frac{2}{C-iGS}\right)~,
\end{equation}
with $G=R^{[0]}K_{z}^{[1]}/R^{[1]}K_{z}^{[0]}$.

Recall that it was (and continues to be) assumed that both $R^{[0]}$ and $C^{[0]}$ are real and non-dispersive. The case (e.g., such as in a macroscopically-homogeneous porous layer) in which either or both $R^{[1]}$ and $C^{[1]}$  are complex and dispersive demands a very involved analysis that is out of the scope of the present discussion, so that we shall make the drastically-simplified assumption that both $R^{[1]}$ and $C^{[1]}$  are real and non-dispersive. We define the modulus squared of the top transfer function as
\begin{equation}\label{7-320}
T(f)=\left\|
\frac{u^{[1]}(x,H,f)}
{A^{[0]+}(f)}
\right\|^{2}=\left\|
\frac{A^{[1]}(f)}
{A^{[0]+}(f)}
\right\|^{2},
\end{equation}
so that on account of our drastic assumptions
\begin{equation}\label{7-330}
T(f)=\frac{4}{C^{2}+G^{2}S^{2}}.
\end{equation}
By equating to zero the partial derivative of this expression with respect to $f$ and, on account of the fact that $G$ is now independent of $f$, we obtain
\begin{equation}\label{7-340}
(1-G^{2})CS=0~,
\end{equation}
which means that $T$ is extremal when either $C=0$ or $S=0$. To see which of these roots correspond to maximal $T$, we use
\begin{equation}\label{7-350}
T(f)\Big|_{S=0}=4~~,~~T(f)\Big|_{C=0}=\frac{4}{G^{2}}~,
\end{equation}
so that the identification of the frequencies at which $T$ is maximal relies on whether $G>1$ or $G<1$. It is straightforward to show that
\begin{equation}\label{7-350}
G=\frac{R^{[0]}C^{[0]}\eta^{[0]}}{R^{[1]}C^{[1]}\eta^{[1]}}~,
\end{equation}
wherein $\eta^{[l]}=\sqrt{1-\left(\frac{C^{[0]}\sin\theta^{i}}{C^{[l]}}\right)^{2}}$. For small $|\theta^{i}|$ and reasonable velocity contrasts, $\eta^{[l]}\approx 1$, so that
\begin{equation}\label{7-360}
G\approx\frac{R^{[0]}C^{[0]}}{R^{[1]}C^{[1]}}~.
\end{equation}
If we refer to the porous material of fig. \ref{gdd11params}, neglect the dispersion and imaginary parts of the effective wavespeed and density, then a plausible choice of the real parts is: $R^{[l]}=2~Kg~ m^{-3}$ and $C^{[l]}=250~ms^{-1}$, considering, as previously, that
 $R^{[0]}=1.2~Kg~ m^{-3}$ and $C^{[0]}=342~ms^{-1}$, so that $G\approx 0.821$ which fact shows that $C=0$ corresponds to a maximum of $T$ and $S=0$ to a minimum of $T$. Moreover, $C=0$ implies that $K_{z}^{[1]}H=(2m+1)\pi/2$, or $K^{[1]}H\approx(2m+1)\pi/2$ for $l=0,1,...$. This means that the response function $T$ is maximal for layer thicknesses
\begin{equation}\label{7-370}
H_{m}\approx\frac{C^{[1]}}{4f}(2m+1)=\frac{\Lambda^{[1]}}{4}(2m+1)~;~m=0,1,....~.
\end{equation}
or for frequencies
\begin{equation}\label{7-380}
f_{m}\approx\frac{C^{[1]}}{4H}(2m+1)~;~m=0,1,....~.
\end{equation}

Thus, from the acoustic point of view, the top response function $T$ is maximal, but not generally infinite, for layer thickness that are odd multiples of the quarter wavelength $\Lambda^{[1]}$, a finding we could qualify as 'quarter wavelength maxima' (QWM) or 'quarter wavelength pseudo resonance' (QWPR), rather than QWR. However, it should be noted that these maxima tend to infinity as $G$ tends to zero, which means that QWM tends to QWR as $G\rightarrow 0$.

On the other hand, from the elastic wave point of view, the top response function $T$ is maximal, for frequencies $f_{m}$ that are odd multiples of $\frac{C^{[1]}}{4H}$ and $f_{0}=\frac{C^{[1]}}{4H}$  is named the 'soil (layer) natural frequency' or simply the 'soil frequency'. Once again, as $G\rightarrow 0$, the maximum of response tends to infinity which fact is reminiscent of a resonance so that $f_{0}$ is also named the 'soil layer resonant frequency' or '1D resonance frequency' by geophysicists.

The adjective 'natural' in connection with a frequency suggests that some sort of mode is involved as concerns the field within the layer. Recall that the boundary condition on the top $z=H$ was $U_{,z}(x,H,f)=0$. Eq. (\ref{7-310}) shows that
\begin{equation}\label{7-390}
U^{[1]}(x,0,f)=A^{[1]}C\exp(iK_{x}x)~,
\end{equation}
so that
\begin{equation}\label{7-400}
U^{[1]}(x,0,f_{m})=0~,
\end{equation}
which means that at the frequencies $f_{m}$, the layer behaves like a cavity whose upper face is the locus of a Neumann (rigid) boundary condition and whose lower face is the locus of a Dirichlet (pressure release) boundary condition. Such a situation suggests that the field in  the layer is a mode of the cavity at the frequencies $f_{m}$ and therefore it is legitimate to term the latter the 'natural (or eigen) frequencies' of the  cavity modes.

Another consequence of (\ref{7-380}) is that the first maximum of top response shifts to lower frequencies as $H$ increases or $C^{[1]}$ decreases, and a final consequence of this formula is that the $f_{m}$ do not depend on $G$. In fact $G$ only controls the height of the maxima of the response function.

All these features turn out to hold remarkably well for  rigidly-backed layers (see fig. \ref{zerozb 01} in sect. \ref{numefflayer}) even when the drastic assumption of real, non-dispersive effective wavespeed and density is relaxed as is necessary when considering a foam material with characteristics such as is depicted in fig. \ref{gdd11params}. Moreover, since the top $T$ is related to $\|A^{[1]}\|^{2}$, as is the absorbed flux in the layer, many of the above-mentioned features of $T$ apply to, and help to understand, the layer absorption too. As we shall see further on in sect. \ref{numefflayer}, these features hold rather well at low frequencies even when the layer is replaced by a grating of alternating (in the $x$ direction) blocks of rigid material and foam material, this being the real reason why the QWR (more properly the QWM) plays a useful role in the understanding of how such a structure responds (notably as concerns  absorption) to an airborne acoustic plane wave.
\section{Analytical aspects of the $M=0$ approximation of the response of the grating structure}\label{Meq0}\label{Meq0}
A consequence of  (\ref{5-010}) is
\begin{equation}\label{8-020}
Y_{0}^{(0)}=\frac{Z_{0}}{X_{00}^{(0)}}~,
\end{equation}
whence
\begin{equation}\label{8-025}
Y_{0}^{(0)}=a_{0}^{[1](0)}=a^{[0]+}\left[\frac{E_{00}^{+}}{\cos\left(k_{z0}^{[1]}h\right)+
\frac{w}{4id}\frac{\rho^{[0]}}{\rho^{[1]}}k_{z0}^{[1]}\sin\left(k_{z0}^{[1]}h\right)\Sigma_{00}^{(0)}}\right]
~.
\end{equation}
However,
\begin{equation}\label{8-027}
E_{00}^{\pm}=2\text {sinc}(k_{x0}^{[0]}w/2)~~,~~
\Sigma_{00}^{(0)}=\frac{1}{k_{z0}^{[0]}}E_{00}^{+}E_{00}^{-}=
\frac{4}{k_{z0}^{[0]}}[\text {sinc}( k_{x0}^{[0]}w/2)]^{2}
~,
\end{equation}
so that
\begin{equation}\label{8-030}
a_{0}^{[1](0)}=a^{[0]+}\left[
\frac{2\text {sinc}\big(k_{x0}^{[0]}w/2\big)}
{
\cos\big(k_{z0}^{[1]}h\big)+
\frac{w}{id}\frac{\rho^{[0]}}{\rho^{[1]}}\frac{k_{z0}^{[1]}}{k_{z0}^{[0]}}\Big(\text {sinc}\big(k_{x0}^{[0]}w/2\big)\Big)^{2}
\sin\big(k_{z0}^{[1]}h\big)
}
\right]
~.
\end{equation}
Moreover
\begin{equation}\label{8-035}
a_{0}^{[0]-(0)}=a_{0}^{[0]+}-a_{0}^{[1](0)}\frac{w}{2id}\frac{\rho^{[0]}}{\rho^{[1]}}
\frac{k_{z0}^{[1]}}{k_{z0}^{[0]}}\sin\big(k_{z0}^{[1]}h\big)E_{00}^{+}
~,
\end{equation}
which reduces to
\begin{equation}\label{8-040}
a_{0}^{[0]-(0)}=a^{[0]+}\left[
\frac{
\cos\big(k_{z0}^{[1]}h\big)
-\frac{w}{id}\frac{\rho^{[0]}}{\rho^{[1]}}\frac{k_{z0}^{[1]}}{k_{z0}^{[0]}}
\Big(\text {sinc}\big(k_{x0}^{[0]}w/2\big)\Big)^{2}\sin\left(k_{z0}^{[1]}h\right)}
{\cos\big(k_{z0}^{[1]}h\big)+\frac{w}{id}\frac{\rho^{[0]}}{\rho^{[1]}}\frac{k_{z0}^{[1]}}{k_{z0}^{[0]}}
\Big(\text {sinc}\big(k_{x0}^{[0]}w/2\big)\Big)^{2}\sin\big(k_{z0}^{[1]}h\big)}
\right]
~.
\end{equation}
These results call for three comments. The first  is that there is a striking resemblance of these amplitudes with the layer configuration amplitudes in sect. \ref{layramps}. This similarity will be exploited further on.
The second comment is just to recall that  we can compute the $M=0$ approximation of reflected and absorbed fluxes via (\ref{6-092})-(\ref{6-094}) associated with these amplitudes.
 The third comment has to do with the possibility (or impossibility) of surface-wave resonances (SWR, not to be confounded with the QWR) showing up in the response functions. SWR, typically those associated with the excitation of homogeneous layer modes (typically of the Love variety \cite{ejp57}), occur (i.e., at a  set of frequencies) for which the denominator in the response amplitudes are equal (in the absence of losses in the media in presence) or very nearly equal  (in the presence of slightly-lossy media) to zero, therefore leading to infinite or very large response. For this to occur, while assuming that the medium in the grating groove is lossless (or very slightly lossy), would require that the  second term (i.e., the one including $\sin$)  in the denominators of (\ref{8-030}) and (\ref{8-040}) be real and negative in relation to the first (i.e., the $\cos$) term, but this is impossible because the factor $w/id$ multiplying $\sin$ is imaginary. It follows that the $M=0$ approximation of the uneven boundary response cannot account for SWR behavior, which fact was already observed in the numerical results presented in sect. \ref{M012}. We shall return to this issue in sect. \ref{comp}.
\subsection{From the  $M=0$ approximate solution to an effective medium representation of the response of the grating configuration} \label{comp}
Let
\begin{equation}\label{8-050}
\mathfrak{S}=\frac{\sin\big(k_{z0}^{[1]}h\big)}{\text {sinc}\big(k_{x0}^{[0]}w/2\big)}~~,~~\mathfrak{C}=\frac{\cos\big(k_{z0}^{[1]}h\big)}{\text {sinc}\big(k_{x0}^{[0]}w/2\big)}
~.
\end{equation}
Then
\begin{equation}\label{8-060}
a_{0}^{[1](0)}=a^{[0]+}\left[
\frac{2}
{
\mathfrak{C}+
\frac{w}{id}\frac{\rho^{[0]}}{\rho^{[1]}}\frac{k_{z0}^{[1]}}{k_{z0}^{[0]}}\Big(\text {sinc}\big(k_{x0}^{[0]}w/2\big)\Big)^{2}
\mathfrak{S}\Big)
}
\right]
~.
\end{equation}
\begin{equation}\label{8-070}
a_{0}^{[0]-(0)}=a^{[0]+}\left[
\frac{\mathfrak{C}-
\frac{w}{id}\frac{\rho^{[0]}}{\rho^{[1]}}\frac{k_{z0}^{[1]}}{k_{z0}^{[0]}}\Big(\text {sinc}\big(k_{x0}^{[0]}w/2\big)\Big)^{2}
\mathfrak{S}\Big)}
{
\mathfrak{C}+
\frac{w}{id}\frac{\rho^{[0]}}{\rho^{[1]}}\frac{k_{z0}^{[1]}}{k_{z0}^{[0]}}\Big(\text {sinc}\big(k_{x0}^{[0]}w/2\big)\Big)^{2}
\mathfrak{S}\Big)
}
\right]
~.
\end{equation}
The comparison of (\ref{8-060})-(\ref{8-070}) with (\ref{7-190})-(\ref{7-195}) shows that the zeroth-order approximate solution of the grating  problem is structurally-similar to the exact solution of the homogeneous layer problem. This suggests that there exists a relation of the homogeneous layer parameters H,$C^{[1]},R^{[1]}$ to the grating parameters $h,c^{[1]},\rho^{[1]}$. To establish this relation is equivalent to solving an inverse problem, and since it is well-known that the solution of inverse problems are generally not unique, we can expect to be able to find many possible relations of H,$C^{[1]},R^{[1]}$ to  $h,c^{[1]},\rho^{[1]}$.
\subsubsection{General considerations on inverse problems such as ours}
There exists a host of manners of formulating the inverse parameter retrieval problem. Basically, one tries to minimize some function of the discrepancy between the 'data' involving so-called 'true parameters' and a hypothetical model that one thinks is able to generate the data, this model involving the so-called 'trial  parameters'. Note that this minimization problem has to be solved for many frequencies (i.e., at least those in the bandwidth of the acoustic solicitation) since the data is frequency-dependent even if the true parameters do not depend on frequency.

In the present case, the true parameters form the set $\mathbf{p}=\{h,c^{[1]},\rho^{[1]}\}$ and the trial parameters form the set $\mathbf{P}= \{\text{H},C^{[1]},R^{[1]}\}$. The response function data is $a_{0}^{[1](0)}$, but it could be $a_{0}^{[0]-(0)}$ or both of these. The data is more traditionally a measurable quantity such as the pressure at one or several points within or outside of the grating, these pressures being, of course directly related to $a_{0}^{[1](0)}$ and $a_{0}^{[0]-(0)}$.

Let a response function associated with the true parameters be $\mathfrak{f}(\mathbf{p},\omega)$ and its counterpart associated with the trial parameters be $\mathfrak{F}(\mathbf{P},\omega)$. Since we reserve the possibility of employing several response functions to deal with our inverse problems, we attach a subscript $j$ to $\mathfrak{f}$ and $\mathfrak{F}$, with $j$ ranging from 1 to $N_{d}$, the latter quantity being the number of response functions we deal with. Often, $N_{d}$ is taken equal to the number of unknown parameters in $P$.

If we are very sure of the veracity of our trial model then we can attempt to solve for $\mathbf{P}$ via the system of equations
\begin{equation}\label{8-080}
\mathfrak{F}_{j}(\mathbf{P}(\omega),\omega)-\mathfrak{f}_{j}(\mathbf{p}(\omega),\omega)=0
~j=1,2,...,N_{d}~.
\end{equation}
Note that we have explicitly introduced the frequency-dependence of the two parameter sets, since this dependence reflects the reality of physical problems such as the one we are faced with. This, and the previously-mentioned remark, mean that the set of (generally non-linear) equations (\ref{8-080}) must be solved for all the frequencies within the bandwidth of the source.

If we are less sure of the veracity of the trial model, or unable to solve the set of nonlinear equations explicitly, then a common strategy is to to search for $\mathbf{P}$ in the following optimization problem manner
\begin{equation}\label{8-090}
\mathbf{\tilde{P}}(\omega)=\arg\min _{\mathbf{P}\in\mathcal{S}}\sum_{j=1}^{N_{d}}\|\mathfrak{F}_{j}(\mathbf{P}(\omega),\omega)-\mathfrak{f}_{j}(\mathbf{p}(\omega),\omega)\|^{2}
~,
\end{equation}
which means that, for each $\omega$, $\mathbf{P}$ is varied over the multidimensional (dimension $L$ equal to the number of to-be-retrieved parameters) search domain, the cost function $\sum_{j=1}^{N_{d}}\|\mathfrak{F}_{j}(\mathbf{P}(\omega),\omega)-\mathfrak{f}_{j}(\mathbf{p}(\omega),\omega)\|^{2}$ is computed for each trial  $\mathbf{P}$, and the so-called optimal  $\mathbf{P}$, denoted by $\tilde{\mathbf{P}}$ is chosen to be the one giving rise to the minimal cost. The so-called effective or optimal model is then $\tilde{\mathfrak{F}}_{j}(\omega)=\mathfrak{F}_{j}(\mathbf{\tilde{P}}(\omega),\omega)$. This procedure, like the one associated with solving a system of nonlinear equations, usually does not give rise to a unique solution. Moreover, the minimization procedure may be unstable, which means that the slightest modification of $\mathfrak{F}$ or $\mathfrak{f}$ can lead to entirely-different $\mathbf{\tilde{P}}$.

It can be advantageous for  the optimization to be carried out for a number  of unknowns in $\mathbf{P}$ inferior to $L$ provided, of course, that one has a decent idea of the values to be assigned to the other parameters in $\mathbf{P}$. This advantage stems from the facts: i) as concerns the procedure (\ref{8-090}), that the non-uniqueness and instability decrease with the dimensionality of the search space, and ii) as concerns (\ref{8-080}), that it might be possible to solve explicitly  the system of equations if there are less unknowns, therefore affording physical insights that are otherwise difficult to obtain from numerical solutions.
\subsection{The methods of solution of the inverse problem adopted herein}\label{3meth}
Hereafter, we treat the inverse problem via (\ref{8-080}) in which $N_{d}=2$,
\begin{equation}\label{8-100}
\begin{array}{c}
\mathfrak{f}_{1}(\omega)=a_{0}^{[1](0)}(\omega)=a^{[0]+}\left[\frac{2\text{sinc}(k_{x0}w/2)}
{c+\frac{w}{id}\frac{\rho^{[0]}}{\rho^{[1]}}\frac{k_{z0}^{[1]}}{k_{z0}^{[0]}}\Big(\text {sinc}\big(k_{x0}^{[0]}w/2\big)\Big)^{2}s\Big)}
\right]~,\\
\mathfrak{F}_{1}(\omega)=A_{0}^{[1]}(\omega)=A^{[0]+}\left[\frac{2}
{C+\frac{R^{[0]}}{iR^{[1]}}\frac{K_{z}^{[1]}}{K_{z}^{[0]}}S}
\right]~,\\
\mathfrak{f}_{2}(\omega)=a_{0}^{[0]-(0)}(\omega)=a^{[0]+}\left[\frac{c-\frac{w}{id}\frac{\rho^{[0]}}{\rho^{[1]}}\frac{k_{z0}^{[1]}}{k_{z0}^{[0]}}\Big(\text {sinc}\big(k_{x0}^{[0]}w/2\big)\Big)^{2}s\Big)}
{c+\frac{w}{id}\frac{\rho^{[0]}}{\rho^{[1]}}\frac{k_{z0}^{[1]}}{k_{z0}^{[0]}}\Big(\text {sinc}\big(k_{x0}^{[0]}w/2\big)\Big)^{2}s\Big)}
\right]~,\\
\mathfrak{F}_{2}(\omega)=A_{0}^{[0]-}(\omega)=A^{[0]+}\left[\frac{C-\frac{R^{[0]}}{R^{[1]}}\frac{K_{z}^{[1]}}{K_{z0}^{[0]}}S}
{C+\frac{R^{[0]}}{iR^{[1]}}\frac{K_{z}^{[1]}}{K_{z}^{[0]}}S}
\right]~,
\end{array}
\end{equation}
wherein
\begin{equation}\label{8-105}
C=\cos(K_{z}^{[1]}\text{H})~,~~C=\sin(K_{z}^{[1]}\text{H})~~,~~c=\cos(k_{z0}^{[1]}h)~~,~~s=\sin(k_{z0}^{[1]}h)~.
\end{equation}
and we solve for only two  of the trial parameters H,$C^{[1]},R^{[1]}$ of the homogeneous layer model at a time (this is similar to the NRW technique of parameter retrieval \cite{qk11} popular in metamaterial research).
 Thus, we search  these two unknowns from
\begin{equation}\label{8-115}
\begin{array}{c}
A^{[1]}(\omega)-a_{0}^{[1](0)}(\omega)=0~,\\
A^{[0]-}(\omega)-a_{0}^{[0]-(0)}(\omega)=0
 \end{array}~.
\end{equation}
It turns out to be difficult to solve this system for two of the parameters H,$C^{[1]},R^{[1]}$ in explicit manner without making the assumption (justified, in that we are essentially interested in the low frequency context)  that the frequency and incident angle are such that
\begin{equation}\label{8-120}
k_{x0}^{[0]}w/2<<1~,
\end{equation}
which implies $\text{sinc}(k_{x0}w/2)\approx 1$ (but not necessarily normal incidence), and therefore
\begin{equation}\label{8-125}
\begin{array}{c}
\mathfrak{f}_{1}(\omega)=a_{0}^{[1](0)}(\omega)\approx \check{a}_{0}^{[1]}(\omega)=a^{[0]+}\left[\frac{2}
{c+\frac{w}{id}\frac{\rho^{[0]}}{\rho^{[1]}}\frac{k_{z0}^{[1]}}{k_{z0}^{[0]}}s}
\right]~,\\
\mathfrak{f}_{2}(\omega)=a_{0}^{[0]-(0)}(\omega)\approx \check{a}_{0}^{[0]-}(\omega)=a^{[0]+}\left[\frac{c-\frac{w}{id}\frac{\rho^{[0]}}{\rho^{[1]}}\frac{k_{z0}^{[1]}}{k_{z0}^{[0]}}s}
{c+\frac{w}{id}\frac{\rho^{[0]}}{\rho^{[1]}}\frac{k_{z0}^{[1]}}{k_{z0}^{[0]}}s}
\right]~,\\
\end{array}
\end{equation}
 It seems natural to: i)  excite the two (true and trial) configurations with the same wave (i.e., the  angles of incidence are the same, as assumed from the outset, and the spectral amplitudes of the incident plane waves are the same (i.e., $A^{[0]+}=a^{[0]+}$), and ii) assume that the media in which this plane wave propagates to be the same (i.e., air). Consequently,   $R^{[0]}=\rho^{[0]}$ and $C^{[0]}=c^{[0]}$ (which implies $K_{z}^{[0]}=k_{z0}^{[0]}$), so that
\begin{equation}\label{8-120}
\begin{array}{c}
\mathfrak{F}_{1}(\omega))=
a^{[0]+}\left[
\frac{2}{C+\frac{\rho^{[0]}}{iR^{[1]}}\frac{K_{z}^{[1]}}{k_{z0}^{[0]}}S}
\right],\\\\
\mathfrak{F}_{2}(\omega)=
a^{[0]+}\left[
\frac{C-\frac{\rho^{[0]}}{R^{[1]}}\frac{K_{z}^{[1]}}{k_{z0}^{[0]}}S}
{C+\frac{\rho^{[0]}}{iR^{[1]}}\frac{K_{z}^{[1]}}{k_{z0}^{[0]}}S}
\right]
\end{array}~.
\end{equation}
Therefore (\ref{8-115}) is of the form
\begin{equation}\label{8-140}
\begin{array}{c}
\left[\frac{2}
{C+\frac{\rho^{[0]}}{iR^{[1]}}\frac{K_{z}^{[1]}}{k_{z0}^{[0]}}S}
\right]-
\left[\frac{2}
{c+\frac{w}{id}\frac{\rho^{[0]}}{\rho^{[1]}}\frac{k_{z0}^{[1]}}{k_{z0}^{[0]}}s}
\right]=0,\\\\
\left[\frac{C-\frac{\rho^{[0]}}{iR^{[1]}}\frac{K_{z}^{[1]}}{k_{z0}^{[0]}}S}
{C+\frac{\rho^{[0]}}{iR^{[1]}}\frac{K_{z}^{[1]}}{k_{z0}^{[0]}}S}
\right]-
\left[\frac{c-\frac{w}{id}\frac{\rho^{[0]}}{\rho^{[1]}}\frac{k_{z0}^{[1]}}{k_{z0}^{[0]}}}
{c+\frac{w}{id}\frac{\rho^{[0]}}{\rho^{[1]}}\frac{k_{z0}^{[1]}}{k_{z0}^{[0]}}s}
\right]=0
\end{array}~.
\end{equation}
We showed previously that $\left[\frac{2}
{C+\frac{\rho^{[0]}}{iR^{[1]}}\frac{K_{z}^{[1]}}{k_{z0}^{[0]}}S}
\right]$ implies $\left[\frac{C-\frac{\rho^{[0]}}{iR^{[1]}}\frac{K_{z}^{[1]}}{k_{z0}^{[0]}}S}
{C+\frac{\rho^{[0]}}{iR^{[1]}}\frac{K_{z}^{[1]}}{k_{z0}^{[0]}}S}
\right]$ and $\left[\frac{2}
{c+\frac{w}{id}\frac{\rho^{[0]}}{\rho^{[1]}}\frac{k_{z0}^{[1]}}{k_{z0}^{[0]}}s}
\right]$ implies  $\left[\frac{c-\frac{w}{id}\frac{\rho^{[0]}}{\rho^{[1]}}\frac{k_{z0}^{[1]}}{k_{z0}^{[0]}}s}
{c+\frac{w}{id}\frac{\rho^{[0]}}{\rho^{[1]}}\frac{k_{z0}^{[1]}}{k_{z0}^{[0]}}s}
\right]$ so that the two equations in (\ref{8-140}) are equivalent and it suffices to choose one of them, i.e.,
\begin{equation}\label{8-200}
\left[\frac{1}
{C+\frac{\rho^{[0]}}{iR^{[1]}}\frac{K_{z}^{[1]}}{k_{z0}^{[0]}}S}
\right]-
\left[\frac{1}
{c+\frac{w}{id}\frac{\rho^{[0]}}{\rho^{[1]}}\frac{k_{z0}^{[1]}}{k_{z0}^{[0]}}s}
\right]=0~,
\end{equation}
which should enable the retrieval of only one of the parameters H,$C^{[1]},R^{[1]}$, assuming, of course, 'plausible' values for the other two.
The translation of this  'equivalence' is a series of relations between the parameters of the layer and their counterparts of the grating. We enumerate hereafter several possible solutions (for a single parameter) of (\ref{8-200}).
\subsubsection{First solution}\label{first}
 Eq. (\ref{8-200}) is satisfied  provided:
\begin{equation}\label{8-210}
\begin{array}{l}
1a)~\tilde{C}^{[1]}=c^{[1]}~\Rightarrow~\tilde{K}_{z}^{[1]}=k_{z0}^{[1]},\\
1b)~\tilde{\text{H}}=h~\Rightarrow~\tilde{C}=c~,~\tilde{S}=s~,\\
1c)~\tilde{R}^{[1]}=\frac{\rho^{[1]}}{\frac{w}{d}}~.
\end{array}
\end{equation}
Note that 1a)-1b) are the 'plausible' guesses and 1c) their consequence, this retrieval being explicit, of simple nature, and unique. Also, note that  the optimal parameters $\tilde{C}^{[1]},\tilde{R}^{[1]}$ of the layer problem depend on the frequency in exactly the same manner as those of the grating problem. Finally, note that the  model of amplitude response in the effective layer is
\begin{equation}\label{8-212}
\tilde{A}^{[1]}(\omega)=a^{[0]+}\left[\frac{2}
{\tilde{C}+\frac{\rho^{[0]}}{i\tilde{R}^{[1]}}\frac{\tilde{K}_{z}^{[1]}}{k_{z0}^{[0]}}\tilde{S}}
\right]=
a^{[0]+}\left[\frac{2}
{c+
\frac{\rho^{[0]}}
{i\frac{\rho^{[1]}}{\frac{w}{d}}}
\frac{k_{z0}^{[1]}}{k_{z0}^{[0]}}
s}
\right]=\check{a}_{0}^{[1]}(\omega)~.
\end{equation}
Associated with this we also have the effective model of response in the air-filled lower half space
\begin{equation}\label{8-214}
\tilde{A}^{[0]-}(\omega)=a^{[0]+}\left[\frac{\tilde{C}-\frac{\rho^{[0]}}{i\tilde{R}^{[1]}}\frac{\tilde{K}_{z}^{[1]}}{k_{z0}^{[0]}}\tilde{S}}
{\tilde{C}+\frac{\rho^{[0]}}{i\tilde{R}^{[1]}}\frac{\tilde{K}_{z}^{[1]}}{k_{z0}^{[0]}}\tilde{S}}
\right]=
a^{[0]+}\left[\frac{c-
\frac{\rho^{[0]}}
{i\frac{\rho^{[1]}}{\frac{w}{d}}}
\frac{k_{z0}^{[1]}}{k_{z0}^{[0]}}
s}
{c+
\frac{\rho^{[0]}}
{i\frac{\rho^{[1]}}{\frac{w}{d}}}
\frac{k_{z0}^{[1]}}{k_{z0}^{[0]}}
s}
\right]=\check{a}_{0}^{[0]-}(\omega)~,
\end{equation}
as well as the effective reflected and absorbed fluxes
\begin{equation}\label{8-216}
\tilde{\mathcal{R}}(\omega)=\left\|\frac{\check{a}_{0}^{[0]-}}{a^{[0]+}}\right\|^{2}=\check{\rho}(\omega)~~,
~~\tilde{\mathcal{A}}(\omega)=\Im\left[
\left\|
\frac{\check{a}_{0}^{[1]}}{a^{[0]+}}
\right\|^{2}\left(\frac{w}{d}\frac{\rho^{[0]}k_{z0}^{[1]}}{\rho^{[1]}k_{z0}^{[0]}}\right)
\left(\cos(k_{z0}^{[1]}h\right)^{*}\sin(k_{z0}^{[1]}h)\right]=\check{\alpha}(\omega)~,
\end{equation}
which satisfy the conservation of flux relation
\begin{equation}\label{8-216}
\tilde{\mathcal{R}}(\omega)+\tilde{\mathcal{A}}(\omega)=\check{\rho}(\omega)+\check{\alpha}(\omega)=1~,
\end{equation}
%
\subsubsection{Second solution}\label{second}
Eq. (\ref{8-200}) is satisfied  provided:
\begin{equation}\label{7-220}
\begin{array}{l}
2a)~\text{H}=h~,\\
2b)~R^{[1]}=\frac{\rho^{[1]}}{\frac{w}{d}}~\\
2c)~K_{z}^{[1]} \text{ solution ($\ne~k_{z0}^{[1]}$) of } \left[\cos\left(K_{z}^{[1]}h\right)+\frac{w}{id}\frac{\rho^{[0]}K_{z}^{[1]}}{\rho^{[1]}K_{z}^{[0]}}\sin\left(K_{z}^{[1]}h\right)\right]-\\
~~~\left[\cos\left(k_{z0}^{[1]}h\right)+\frac{w}{id}\frac{\rho^{[0]}k_{z0}^{[1]}}{\rho^{[1]}k_{z0}^{[0]}}\sin\left(k_{z0}^{[1]}h\right)\right]
=0~\text{and~}C^{[1]}=\omega/
\sqrt{(K_{z}^{[1]})^{2}+(k_{x0}^{[0]})^{2}}~.\\
\end{array}
\end{equation}
Note that 2a)-2b)  are the 'plausible' guesses and 2c) their consequence. Note also  that now the retrieval of $C^{[1]}$ requires solving a nonlinear equation for each frequency and the solution of this equation is not unique. Finally, note that we did not put tildes on the retrieved parameters because we did not make this choice of optimality.
\subsubsection{Third solution}\label{third}
Eq. (\ref{8-200}) is satisfied  provided:
\begin{equation}\label{7-230}
\begin{array}{l}
3a)~C^{[1]}=c^{[1]},\\
3b)~R^{[1]}=\frac{\rho^{[1]}}{\frac{w}{d}}~\\
3c)~\text{H} \text{ solution ($\ne~h$) of } \left[\cos\left(k_{z0}^{[1]}\text{H}\right)+\frac{w}{id}\frac{\rho^{[0]}k_{z0}^{[1]}}{\rho^{[1]}k_{z0}^{[0]}}\sin\left(k_{z0}^{[1]}\text{H}\right)\right]-\\
~~~\left[\cos\left(k_{z0}^{[1]}h\right)+\frac{w}{id}\frac{\rho^{[0]}k_{z0}^{[1]}}{\rho^{[1]}k_{z0}^{[0]}}\sin\left(k_{z0}^{[1]}h\right)\right]=0~.
\end{array}
\end{equation}
Note that 3a)-3b)  are the 'plausible' guesses and 3c) their consequence. Note also  that now the retrieval of $\text{H}$ requires solving a nonlinear equation for each frequency and the solution of this equation is not unique. Finally, note that we did not put tildes on the retrieved parameters because we did not make this choice of optimality.
\subsubsection{Further comments on consequences of the $M=0$ approximation of grating response}\label{comm}
If, for a reason to be evoked further on, one chooses one of the three solutions as a means of identifying some or all of the effective medium parameters (i.e., those denoted by upper-case letters), then he should be aware of the fact that these choices all derive from a $M=0$ approximation of the uneven boundary response, which, as shown previously in sect. \ref{3meth}, is only valid at very low frequencies and/or near-normal incidence,  cannot account, by any means,  for surface wave resonant behavior of the uneven boundary. To deal with SWR behavior, one must solve an inverse problem by matching 'true data' deriving from the $M=1$ approximation to the layer model thereof, as is done in \cite{wi18b}.

From here~ on,~ we shall ~make the first~ choice of effective ~model, i.e., the ~rigidly-backed macroscopically-homogeneous layer whose constitutive and geometric parameters are $C^{[1]}=c^{[1]}$, H=$h$, and $R^{[1]}=\frac{\rho^{[1]}}{\frac{w}{d}}$.
\section{Numerical results for the  foam filler of \cite{gdd11}: comparison of the $w/d<1$ grating quasi-exact and effective-layer response to the foam-filled layer response}\label{numefflayer}
At this point, it seems useful to accomplish two tasks:\\
1) show to what extent  all the functions (such as the absorbed flux $\check{\alpha}(\omega)$) deriving  from $\check{a}_{0}^{[1]}(\omega)$ (itself being a result of  assuming (\ref{8-120}) in  $a_{0}^{[1](0)}(\omega)$) compare with their counterparts deriving from $a_{0}^{[1](0)}(\omega)$,\\
2) show to what extent the grating response, notably the approximation of the absorbed flux $\check{a}_{0}^{[1]}(\omega)$), differs from the rigidly-backed macroscopically-homogeneous layer faced by air response with constitutive and geometric parameters $C^{[1]}=c^{[1]}$, $R^{[1]}=\rho^{[1]}$ and H=$h$, notably the absorbed flux by this layer resulting from $\bar{A}^{[1]}(\omega)$.
\begin{equation}\label{8-270}
\bar{\mathcal{A}}(\omega)=\frac{4\Im(gsc^{*})}{\|c\|^{2}+\|gs\|^{2}+2\Im(gsc^{*})}
~,
\end{equation}
wherein $g=\frac{\rho^{[0]}}{\rho^{[1]}}\frac{k_{z0}^{[1]}}{k_{z0}^{[0]}}$.\\

\begin{figure}[ht]
\begin{center}
\includegraphics[width=0.75\textwidth]{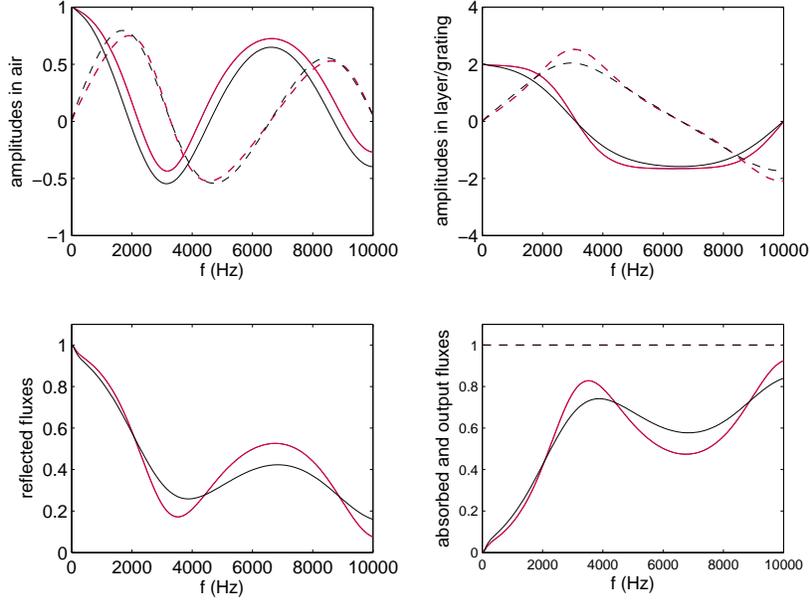}
\caption{The blue curves refer to the $M=0$ approximate responses for the grating=periodically-inhomogeneous layer. The red curves refer to the effective-layer responses of the grating. The black curves refer to the responses of the macroscopically-homogeneous foam-filled layer. Upper left-hand panel: $\Re a_{0}^{[0]-(0)}(\omega)$ (blue full), $\Im a_{0}^{[0]-(0)}(\omega)$ (blue dashed),  $\Re \check{a}_{0}^{[0]-}(\omega)$ (red full), $\Im \check{a}_{0}^{[0]-}(\omega)$ (red dashed referring to the approximate results for the grating),  $\Re \bar{A}^{[0]-}(\omega)$ (black full), $\Im \bar{A}^{[0]-}(\omega)$ (black dashed). Upper right-hand panel: $\Re a_{0}^{[1](0)}(\omega)$ (blue full), $\Im a_{0}^{[1](0)}(\omega)$ (blue dashed),  $\Re \check{a}_{0}^{[1]}(\omega)$ (red full), $\Im \check{a}_{0}^{[1]}(\omega)$ (red dashed),  $\Re \bar{A}^{[1]}(\omega)$ (black full), $\Im \bar{A}^{[1]}(\omega)$ (black dashed). Lower left-hand panel: $\rho^{(0)}(\omega)$ (blue full), $\check{\rho}(\omega)$ (red full) $\bar{\mathcal{R}}(\omega)$ (black full). Lower right-hand panel: $\alpha^{(0)}(\omega)$ (blue full), $\check{\alpha}(\omega)$ (red full) $\bar{\mathcal{A}}(\omega)$ (black full), output fluxes (dashed).  $h=0.02~m$,  $w=0.015~m$, $d=0.02~m$,  $\theta^{i}=0^{\circ}$, $M=0$, and the foam parameters taken (see sect. \ref{gdd}) from \cite{gdd11}.}
\label{zerozb 01}
\end{center}
\end{figure}

\begin{figure}[ptb]
\begin{center}
\includegraphics[width=0.75\textwidth]{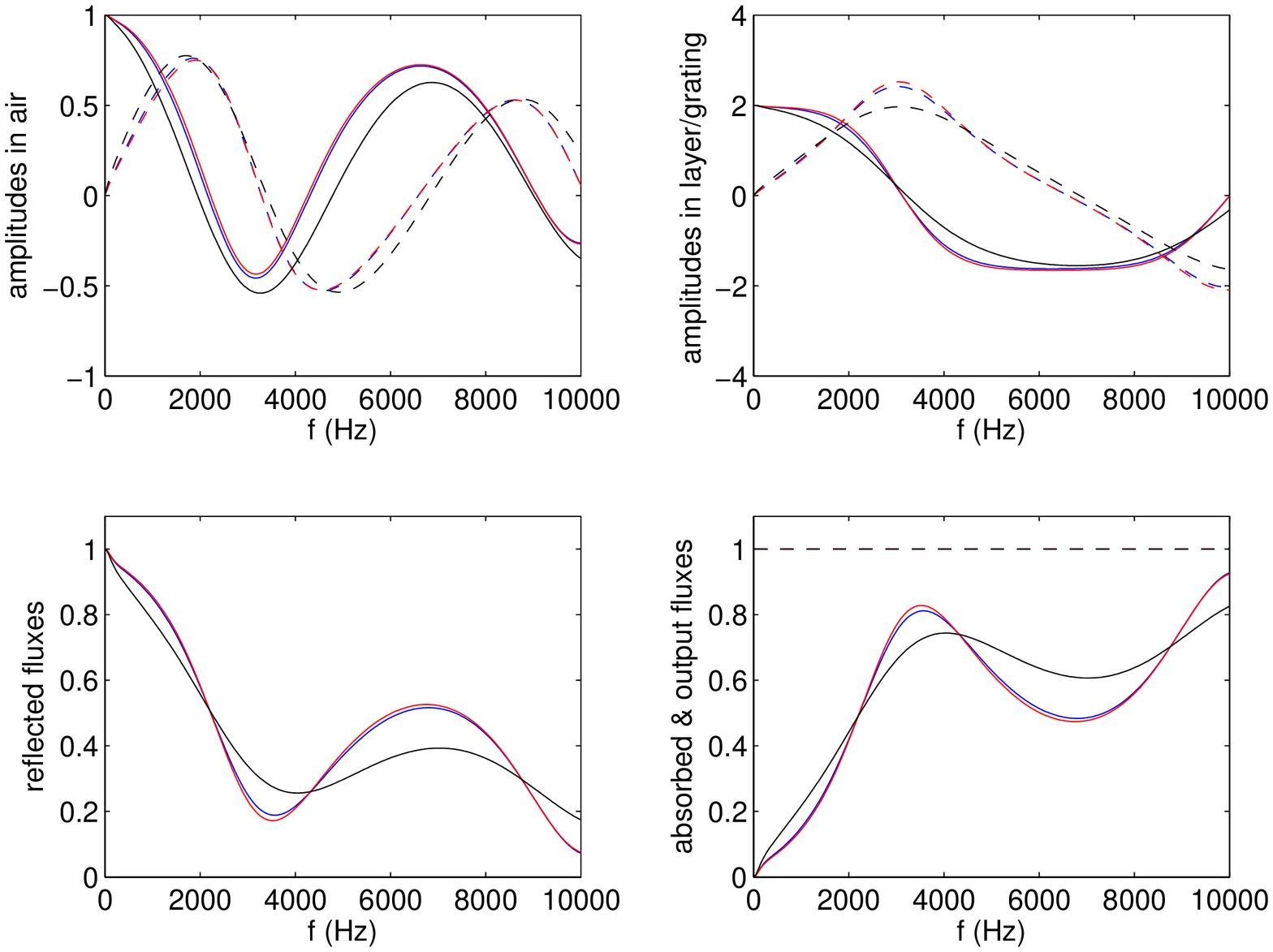}
\caption{Same as fig. \ref{zerozb 01} except that  $\theta^{i}=20^{\circ}$.}
\label{zerozb 02}
\end{center}
\end{figure}

\begin{figure}[ptb]
\begin{center}
\includegraphics[width=0.75\textwidth]{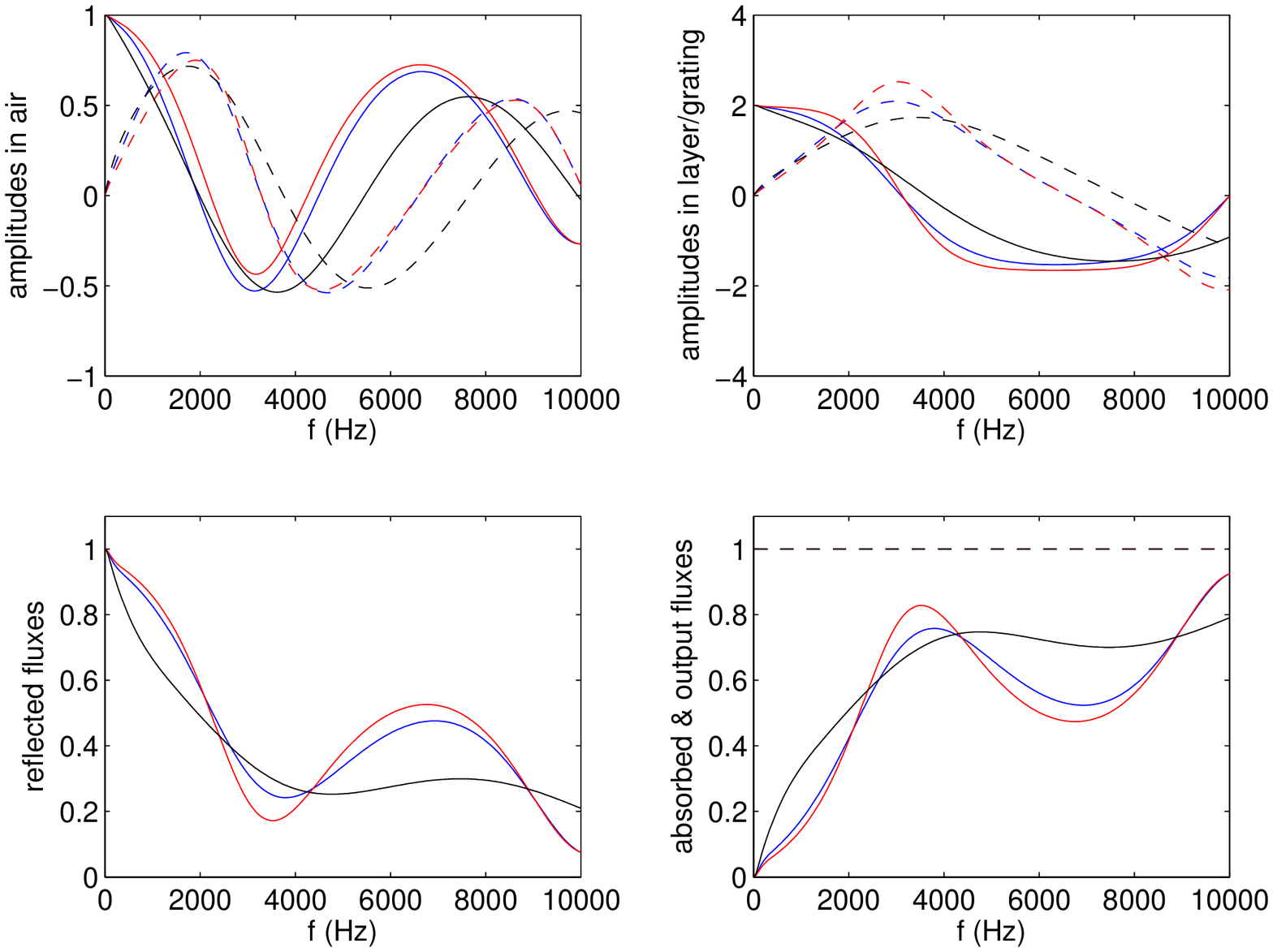}
\caption{Same as fig. \ref{zerozb 01} except that $\theta^{i}=40^{\circ}$.}
\label{zerozb 03}
\end{center}
\end{figure}

\begin{figure}[ptb]
\begin{center}
\includegraphics[width=0.75\textwidth]{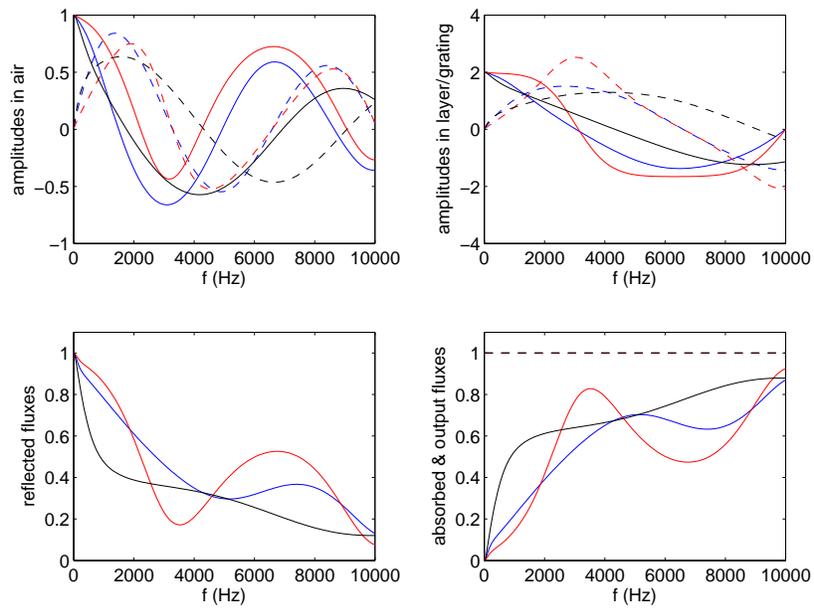}
\caption{Same as fig. \ref{zerozb 01} except that $\theta^{i}=60^{\circ}$.}
\label{zerozb 04}
\end{center}
\end{figure}

In figs. \ref{zerozb 01}-\ref{zerozb 04} the blue curves result from $a_{0}^{[1](0)}(\omega)$, the red curves from $\check{a}_{0}^{[1]}(\omega)$  and the black curves from $\bar{A}^{[1]}(\omega)$. In the upper panels, the full curves denote real parts and the dashed curves imaginary parts of the amplitude functions. The left-hand panels are relative to functions in the lower air half space whereas the right-hand panels are relative to functions in the layer or grating. More specifically: (a) the upper left-hand panel is relative to the reflected amplitudes and the upper right-hand to the  amplitudes in the layer/grating, (b) the lower left-hand panel is relative to the reflected fluxes and the lower right-hand panel to the absorbed fluxes (full curves) and output fluxes (dashed curves).
\clearpage
\newpage
It is seen from these figures that:\\
1. In the low frequency region depicted in the figures, the effective model  responses (red curves) appear to be quite close to the $M=0$ grating responses (the latter having been shown numerically in figs. \ref{m012-05}-\ref{m012-08} to be adequate approximations at low frequencies and small incident angles) for small $|\theta^{i}|$ (up to $\sim 40^{\circ}$) as one would expect from the condition (\ref{8-120}); this authorizes the use of the effective layer model instead of the grating model to predict the low frequency, small incident angle response of the grating.\\
2. The black curves in the upper right-hand panels of figs. \ref{zerozb 01}-\ref{zerozb 02} constitute numerical verifications, for a model of foam with dispersive, complex density and wavespeed, of the predictions of the analysis of sect. \ref{qwr}, even though these predictions were based on the assumption of real, non-dispersive density and wavespeed. In particular,  this analysis predicts quite well: (i) the frequency $f_{0}$ of the lowest-frequency response peak in the layer provided the angle of incidence is near-normal, (ii)   the fact that the real part of the amplitude $A^{[1]}$ in the layer is nil at this frequency and (iii) the fact that the behavior of $A^{[1]}$ is not really resonant near this peak even though the quarter wavelength formula is commonly-thought to be associated with a resonant process.\\
3. The black curves in the upper and lower right-hand panels of figs. \ref{zerozb 01}-\ref{zerozb 02} show that the frequency $f_{0}$ of the pseudo-resonance in $A^{[1]}$ does not coincide with the maximum of absorbed flux, this being due to the fact that the latter depends not only on $\|\frac{A^{[1]}}{A^{[0]+}}\|^{2}$ but also on a slowly-varying  (with respect to $\omega$) factor $E(\omega)$ as per (\ref{7-166}).\\
4. The comparison, relative to the absorption in the lower right-hand panels of figs. \ref{zerozb 01}-\ref{zerozb 02}, of the red-blue curves (grating and effective layer) with the black curves (reference layer absorbed flux designated by  $\bar{\mathcal{A}}$) shows that: (i) not only are the peaks of the red-blue curves shifted to lower frequencies with respect to the black curves, (ii) but also  the red-blue peaks are higher (although narrower) than the black peaks, this meaning that the replacement of the reference layer by the grating enables the sought-for result  of increasing the absorption and producing this increase at lower frequencies. We have not attempted to analytically prove this result (which has also been observed in publications such as \cite{gdd11}), but rather verified numerically its apparently-systematic nature.
\section{Increasing the absorption at low frequencies treated as an optimization problem}
%
\subsection{Analytical approach}
The QWR analysis and numerical examples (not shown here)  indicate that: (i) the thickness $h$ of the layer/grating particularly influences the positions of the response peaks, with the number of these peaks in a frequency interval increasing with $h$, and (ii) the transverse filling factor (of foam relative to all the material in one period $d$, i.e., $\Phi=w/d$) principally influences the height of the absorption peaks. This suggests, for a given thickness $h$ and given foam characteristics, that the optimization (i.e., to attain maximal absorption) be carried out for varying $\Phi$, or, in other words, that the partial derivative of the normalized absorbed flux function with respect to $\Phi$ be equal to zero.

Since it was shown that $\check{\alpha}(\omega)$ accounts quite well for the exact low-frequency response, it is legitimate to adopt $\check{\alpha}(\omega)$ (instead of the quasi-exact grating absorbed flux function which exists only numerically) as a proxy for $\alpha(\omega)$. The interest of doing so is, of course, that we thus dispose of a simple, explicit, algebraic expression for $\check{\alpha}(\omega)$ which can easily be differentiated. Moreover, it was underlined that, by definition, $\tilde{\mathcal{A}}(\omega)=\check{\alpha}(\omega)$, so that, by associating $\tilde{\mathcal{A}}(\omega)$ with the first method of solution of the inverse problem,  varying $\Phi$ amounts to varying the density of the effective layer, since
\begin{equation}\label{9-010}
\tilde{R}^{[1]}=\rho^{[1]}/\Phi~,
\end{equation}
thus affording some physical insight into the absorption optimization problem.

We found previously (\ref{7-198}) that, for a rigidly-backed homogeneous layer of thickness H submitted to an airborne acoustic plane wave:
\begin{equation}\label{9-020}
\mathcal{A}(\omega)=\frac{4\Im(GSC^{*})}{\|C\|^{2}+\|GS\|^{2}+2\Im(GSC^{*})}~,
\end{equation}
so that, on account of (\ref{9-010}) (i.e., the replacement of the homogeneous layer by the effective layer),
\begin{equation}\label{9-030}
\tilde{\mathcal{A}}(\omega)=\frac{4\Phi\Im(g\tilde{S}\tilde{C}^{*})}{\|\tilde{C}\|^{2}+\Phi\|g\tilde{S}\|^{2}+2\Phi\Im(g\tilde{S}\tilde{C}^{*})}=
\frac{4\Phi\Im(gsc^{*})}{\|c\|^{2}+\Phi\|gs\|^{2}+2\Phi\Im(gsc^{*})}=\check{\alpha}(\omega)~,
\end{equation}
in which
\begin{equation}\label{9-040}
g=\frac{\rho^{[0]}}{\rho^{[1]}}\frac{k_{z0}^{[1]}}{k_{z0}^{[0]}}~.
\end{equation}
The partial derivative of $\tilde{\mathcal{A}}(\omega)$ with respect to $\Phi$ translates to
\begin{equation}\label{9-050}
\tilde{\mathcal{A}}_{,\Phi}(\omega)=-\Phi^{2}\|gs\|^{2}+\|c\|^{2}=0~,
\end{equation}
from which we deduce (since $\Phi$ must be positive)
\begin{equation}\label{9-060}
\Phi_{\text{opt}}(\omega)=\frac{\|c\|}{\|gs\|}=
\left\|\frac{\rho^{[1]}k^{[0]}_{z0}}{\rho^{[0]}k_{z0}^{[1]}}\right\|\left\|\cot\left(k_{z0}^{[1]}h\right)\right\|~.
\end{equation}
Previous numerical results show that the characteristics of the foam, $\rho^{[1]}$ and $c^{[1]}$ are slowly-varying functions of frequency except at very low frequencies, so that the factor $\left\|\frac{\rho^{[1]}k^{[0]}_{z0}}{\rho^{[0]}k_{z0}^{[1]}}\right\|$ is a slowly-varying function of $\omega$, which means that $\Phi_{\text{opt}}(\omega)$ is a quasi-periodic function of $\omega$ (other than at very low frequencies) on account of the factor $\cot\left(k_{z0}^{[1]}h\right)$. We shall resort to numerics further on to obtain a more precise picture of the spectral characteristics of $\Phi_{\text{opt}}$.

Another feature of (\ref{9-060}) is that $\|c\|$ can exceed $\|gs\|$ in which case $\Phi_{\text{opt}}(\omega)$ exceeds 1, which is nonsense since the filling fraction is defined by $w/d$ and the width $w$ of the foam-loaded grooves cannot exceed the period $d$. This means that that values of $h$ and/or $\omega$ for which the optimal filling fraction exceeds 1 must be excluded from the optimization analysis, as will be explained in the commentary of the numerical results.

Of great interest is to find out how $\tilde{\mathcal{A}}\Big|_{\Phi=\Phi_{\text{opt}}}$ varies with the various structural parameters and how it compares with the absorbed flux in the reference configuration of a rigidly-backed homogeneous foam layer (with characteristics H=$h$, $\rho^{[1]}$ and $c^{[1]}$), submitted to an airborne plane acoustic wave ($R=^{[0]}=\rho^{[0]}$, $C^{[0]}=c^{[0]}$ and $A^{[0]+}=a^{[0]+}$) previously-designated by $\bar{\mathcal{A}}$. It is easily shown that
\begin{equation}\label{9-070}
\tilde{\mathcal{A}}(\omega)\Big |_{\Phi=\Phi_{\text{opt}}}=\frac{2}{1+\frac{\|gsc\|}{\Im(gsc^{*})}}~.
\end{equation}

For the rigidly-backed homogeneous (i.e., not the effective) layer, (\ref{9-020}) entails
\begin{equation}\label{9-080}
\bar{\mathcal{A}}(\omega)=\frac{4\Im(gsc^{*})}{\|c\|^{2}+\|gs\|^{2}+2\Im(gsc^{*})}=
\frac{2}{
1+\frac{\left(\frac{\|c\|^{2}+\|gs\|^{2}}{2}\right)}
{\Im(gsc^{*})}
}~.
\end{equation}
But  any two complex numbers $Z_{1}$ and $Z_{2}$ satisfy the inequality $\frac{\|Z_{1}\|^{2}+\|Z_{1}\|^{2}}{2}\ge \|Z_{1}Z_{2}\|$, so that
\begin{equation}\label{9-090}
\bar{\mathcal{A}}(\omega)\le\tilde{\mathcal{A}}(\omega)\Big |_{\Phi=\Phi_{\text{opt}}}~;~\forall \omega\ge 0~,
\end{equation}
which means that the optimal absorption in the rigidly-backed  grating structure with alternating foam and rigid regions is larger than or equal to, at all frequencies, the absorption in the rigidly-backed homogeneous foam (the same as that of the grating) layer of thickness equal to that of the grating. This, of course, is the desired feature of using the grating instead of the homogeneous layer.

A last remark concerning (\ref{9-070}): this expression shows that the absorption can be total (i.e. $\tilde{\mathcal{A}}=1$ when
\begin{equation}\label{9-100}
\|gsc\|=\Im(gsc^{*})~.
\end{equation}
Finding the $h,~\omega$ pairs for which this total absorption can occur requires solving the non-linear equation (\ref{9-100}), a task that is not addressed herein.
\subsection{Numerical approach}
%
\subsubsection{Decreasing the frequency for optimal absorption by increasing $h$ for the foam of \cite{gdd11}}
\begin{figure}[ht]
\begin{center}
\includegraphics[width=0.75\textwidth]{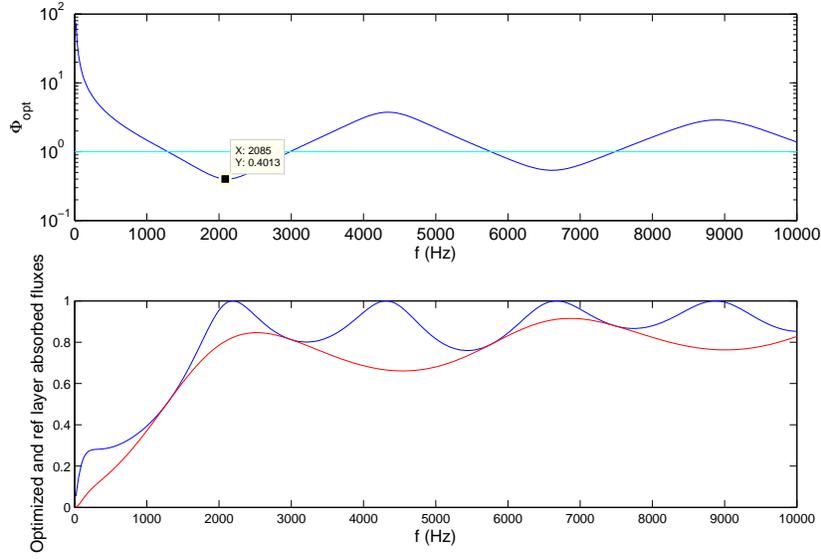}
\caption{$\Phi_{\text{opt}}(f)$ (blue curve in the top panel in which the green line denotes the physical limit of $\Phi$, i.e., $\Phi=1$), $\tilde{\mathcal{A}}(f)\Big |_{\Phi=\Phi_{\text{opt}}}$ (blue curve in the bottom panel) and  $\bar{\mathcal{A}}(f)$ (red curve in the bottom panel). Case $h=0.03~m$,   $\theta^{i}=0^{\circ}$ for the constitutive parameters of the foam of sect. \ref{gdd}.}
\label{optgdd 02}
\end{center}
\end{figure}

\begin{figure}[ptb]
\begin{center}
\includegraphics[width=0.75\textwidth]{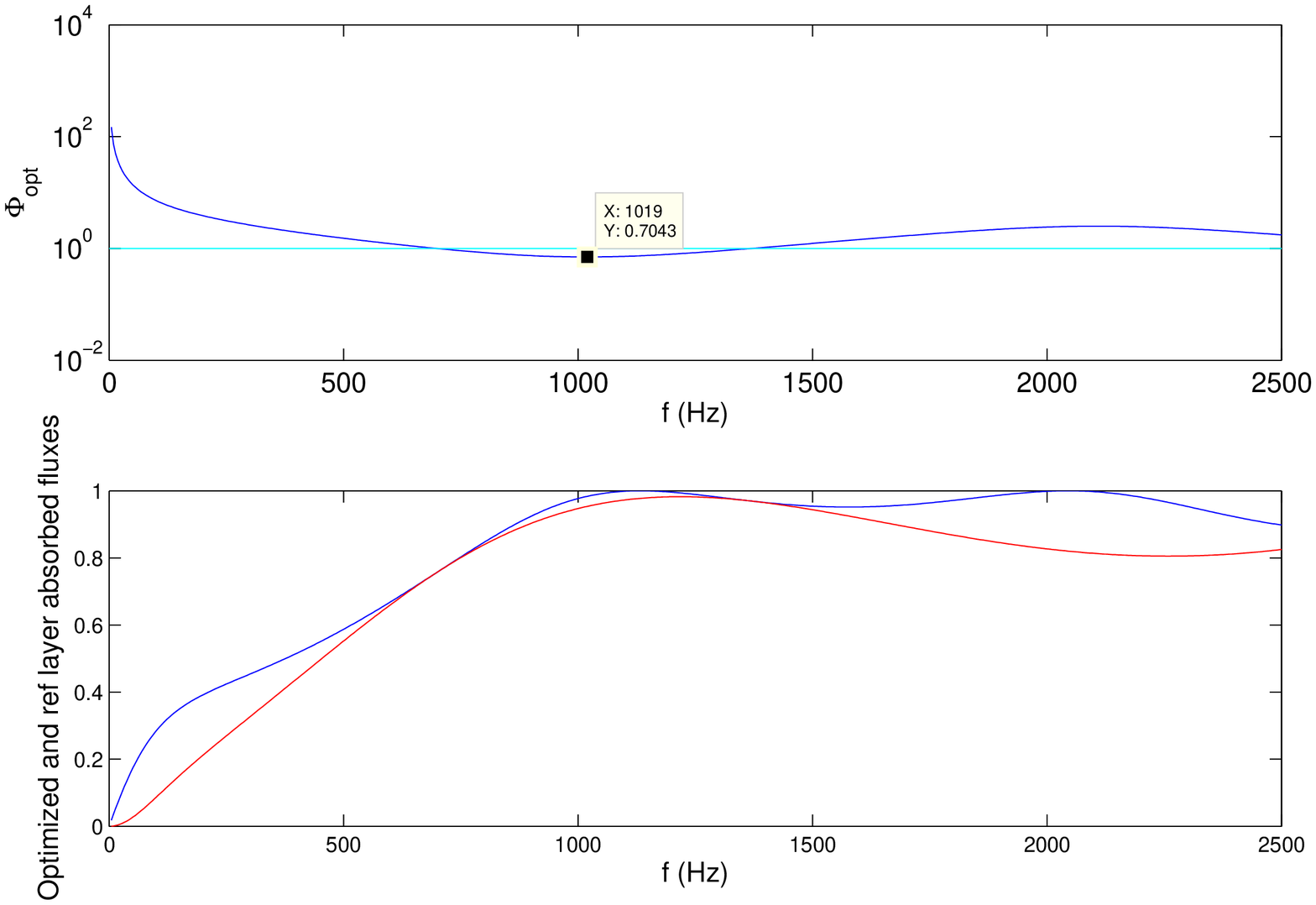}
\caption{Same as fig. \ref{optgdd 02} except that $h=0.06~m$.}
\label{optgdd 03}
\end{center}
\end{figure}

\begin{figure}[ptb]
\begin{center}
\includegraphics[width=0.75\textwidth]{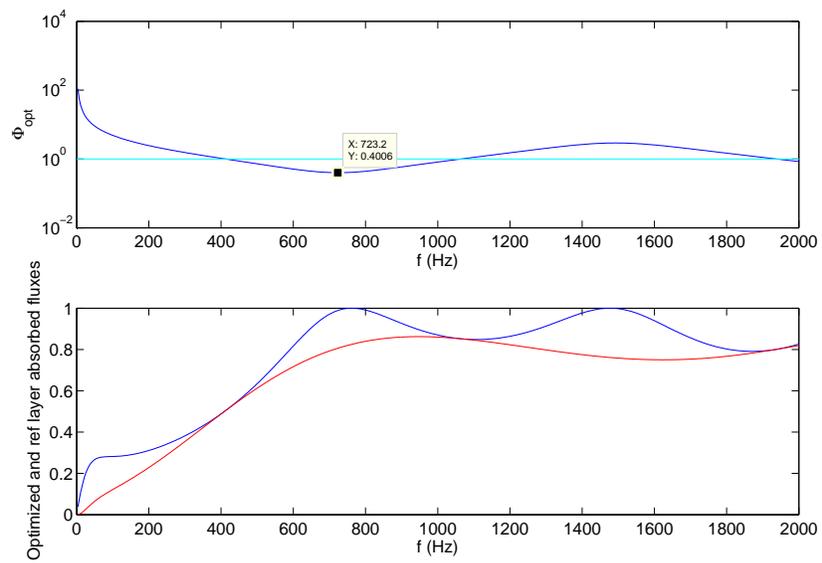}
\caption{Same as fig. \ref{optgdd 02} except that $h=0.1~m$.}
\label{optgdd 04}
\end{center}
\end{figure}

In  figs. \ref{optgdd 02}-\ref{optgdd 04} we plot $\Phi_{\text{opt}}(f)$ (top panels) and $\tilde{\mathcal{A}}(f)\Big |_{\Phi=\Phi_{\text{opt}}}$ as well as $\bar{\mathcal{A}}(f)$ (bottom panels).
\clearpage
\newpage

These results call for the following comments:\\
1. Although the absorption  is optimal for the indicated $h$ and $\omega$, it is not necessarily total (i.e. equal to 1).\\
2. Although the absorption at $\Phi_{\text{opt}}$ is optimal for the indicated $h$ and $\omega$, it is not generally-optimal for other $h$ and $\omega$.\\
3. As observed in the figures, the absorption can be total (i.e., equal to 1) for certain combinations of $\Phi_{\text{opt}},~\omega,~h$.\\
4. $\Phi_{\text{opt}}>1$ is meaningless since the width $w$ of the foam regions of the grating cannot exceed the period $d$. This means that the blue curve in a given bottom panel is meaningful only in the frequency intervals for which the blue curve is below the cyan line in the corresponding upper panel.\\
5. As predicted theoretically, the absorption of the optimized grating (actually its proxy) is superior, at all allowable frequencies, to that of the reference rigidly-backed homogeneous layer (both of thickness $h$) filled with the same foam as that of the grating.\\
6. The position (in terms of frequency) of  optimal absorption $\tilde{\mathcal{A}}(f)\Big |_{\Phi=\Phi_{\text{opt}}}$ does not depend on either $w$ relative to $f$ nor on $d$ relative to $f$, i.e., it depends only on their ratio $w/d$. This is a prediction of a quasi-static model of the grating response and does not necessarily hold for high frequencies (relative to $w$ and $d$). We shall see further on what the effect is of changing $w$ and $d$ while maintaining constant their ratio $\Phi_{\text{opt}}=w/d$.\\
7. The positions (in terms of frequency) of the maxima of the peaks of $\tilde{\mathcal{A}}(f)\Big |_{\Phi=\Phi_{\text{opt}}}$ are shifted to the left (i.e., to lower frequencies) with respect to the maxima of the peaks of $\bar{\mathcal{A}}(f)$. This is the  second desired feature of employing the grating, stressed, and also observed, in (fig. 3 of) \cite{gdd11}, but we have been unable to  mathematically  prove its systematic nature and  establish its amount, this being so because of the dispersive nature of $\rho^{[1]}$ and $c^{[1]}$.
\subsubsection{Decreasing the frequency for optimal absorption by increasing $h$ for the foam of \cite{glv10}}
\begin{figure}[ht]
\begin{center}
\includegraphics[width=0.75\textwidth]{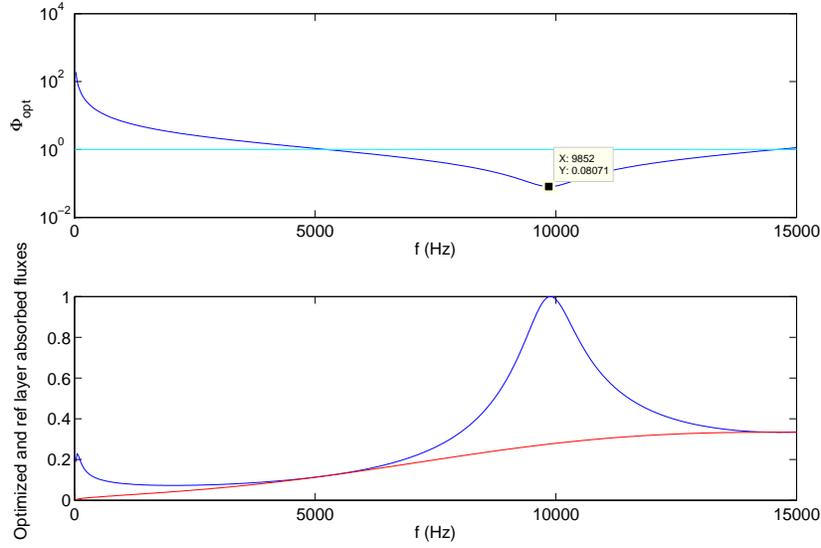}
\caption{$\Phi_{\text{opt}}(f)$ (blue curve in the top panel in which the green line denotes the physical limit of $\Phi$, i.e., $\Phi=1$), $\tilde{\mathcal{A}}(f)\Big |_{\Phi=\Phi_{\text{opt}}}$ (blue curve in the bottom panel) and  $\bar{\mathcal{A}}(f)$ (red curve in the bottom panel). Case $h=0.008~m$,   $\theta^{i}=0^{\circ}$ for the constitutive parameters of the foam of sect. \ref{glv}.}
\label{optglv 01}
\end{center}
\end{figure}

\begin{figure}[ptb]
\begin{center}
\includegraphics[width=0.75\textwidth]{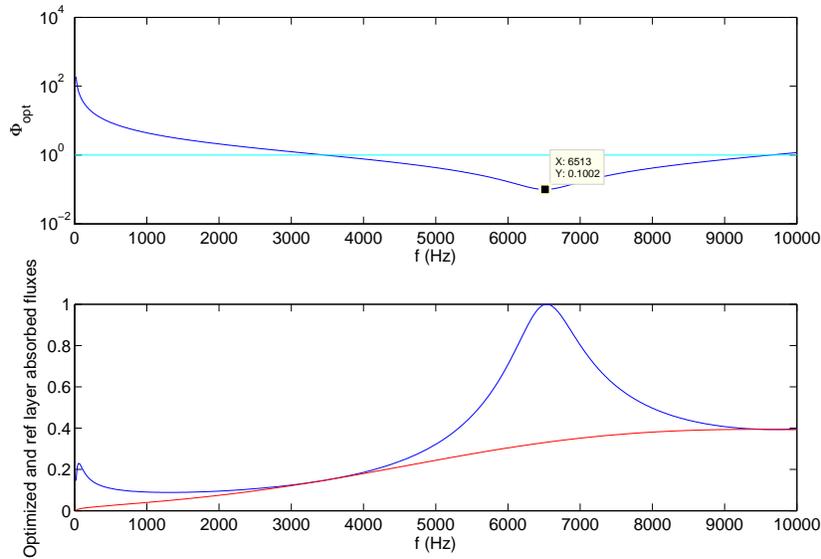}
\caption{Same as fig. \ref{optglv 01} except that $h=0.012~m$,    $\theta^{i}=0^{\circ}$.}
\label{optglv 02}
\end{center}
\end{figure}

The results in \cite{glv10} seem to apply to a rigidly-backed reference layer that is relatively thin (the authors of this publication write that the thickness is $8~mm$, but this figure seems to us to be questionable). Thus, we felt it to be useful to see what happens for relatively-thin foam fillers or layers. This is done in figs. \ref{optglv 01}-\ref{optglv 02}.
\clearpage
\newpage

We thus see that increasing $h$ results in the shift of the the total absorption peak to lower frequencies. Otherwise, these results call for the same comments as in the previous section.
\subsection{Persistence of the enhanced absorption effect for other-than-normal incidence}\label{perth}
Here we start with the optimization parameters of fig. \ref{optgdd 02}, i.e., $\Phi=0.4$, $h=0.03~m$, in a low-frequency regime, which were obtained for normal incidence $\theta^{i}=0^{\circ}$, and inquire as to whether the enhanced (actually total) absorption obtained for normal incidence persists for other angles of incidence as well. This question is of importance in sound absorbiton as well as electromagnetic wave energy harvesting applications.
\begin{figure}[ht]
\begin{center}
\includegraphics[width=0.75\textwidth]{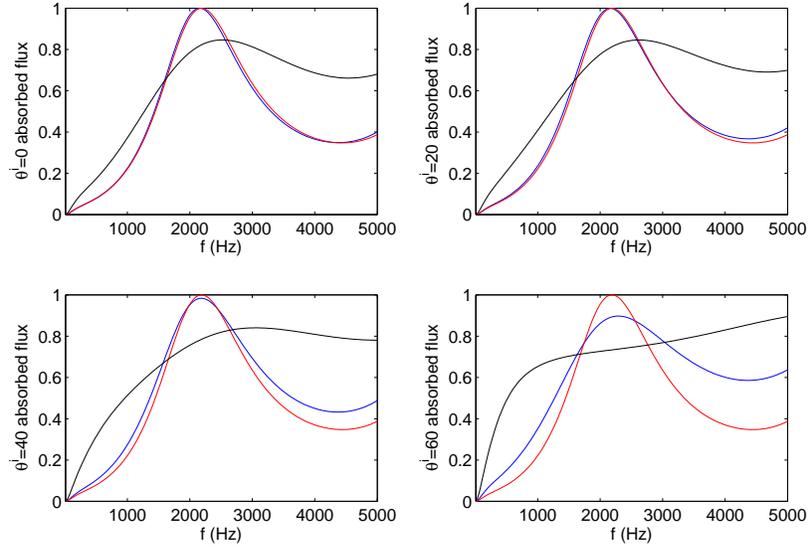}
\caption{In all four panels: $\alpha^{(1)}(\omega)$ (blue full), $\check{\alpha}(\omega)$ (red full) $\bar{\mathcal{A}}(\omega)$ (black full). The upper left-hand, upper right-hand, lower left-hand and lower right-hand panels apply to $\theta^{i}=0,20,40,60^{\circ}$ respectively. The foam parameters are those of sect. \ref{gdd},  $h=0.03~m$, $w=0.004~m$, $d=0.01~m$.}
\label{perth 01}
\end{center}
\end{figure}
\clearpage
\newpage
The answer provided by this figure is clearly affirmative for angles of incidence (in absolute value, and for this choice of grating parameters) that do not exceed $~40^{\circ}$.
\subsection{Numerical results for the $M>0$ grating response concerning the effect of increasing $w$ and $d$ for given $\Phi_{\text{opt}}$ to see if optimal absorption is maintained or changed otherwise}
$\Phi_{\text{opt}}$ was obtained from an analysis employing a proxy of the grating which is an outcome of a low frequency (i.e., $M=0$) approximation. In fact, the frequency does not intervene explicitly in the expression for $\Phi_{\text{opt}}$  nor in that of the corresponding absorbed flux other than through the frequency-dependent properties of the filler material. Thus, it is not at all obvious that the desired increase of absorption (over that of the reference rigidly-backed layer) will be maintained if the frequency is increased beyond static or quasi-static conditions.

The purpose of the graphs in this section is therefore to show how, for a given $\Phi_{\text{opt}}$ and $h$, the absorbed flux of the proxy evolves with increasing $w$ and $d$ while maintaining $w/d$ constant at the value $\Phi_{\text{opt}}$. Since the parameters $w$ and $d$ do not enter in the response of the reference rigidly-backed layer, this response (black curves in the following figures) does not change with $w,~d$. The response of the proxy does depend on $w/d$ but this fraction is constant as a function of $f$ so that the proxy response (red curves) does not change either with $w,~d$. But, of course, the grating response (blue curves) is expected to change with $w,~d$.
\subsubsection{$\Phi_{\text{opt}}=0.7$, $h=0.06~m$, and the foam of \cite{gdd11}}
\begin{figure}[ht]
\begin{center}
\includegraphics[width=0.75\textwidth]{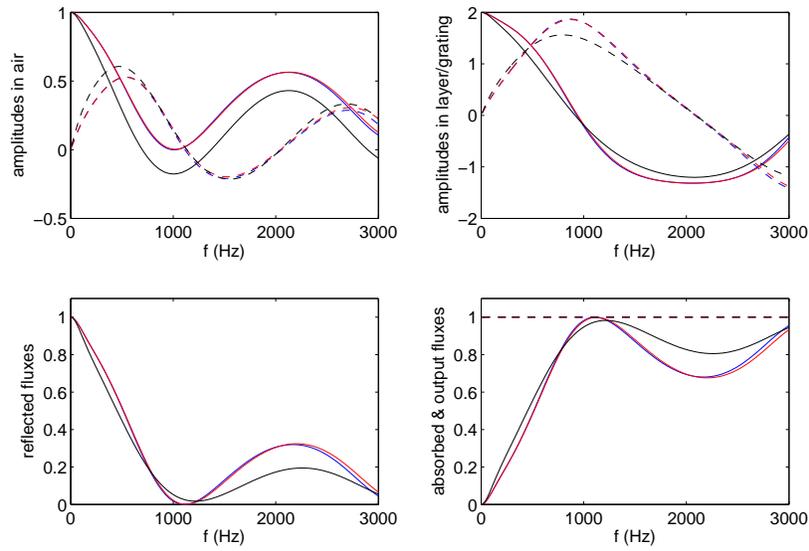}
\caption{Upper left-hand panel: $\Re a_{0}^{[0]-(M)}(\omega)$ (blue full), $\Im a_{0}^{[0]-(M)}(\omega)$ (blue dashed),  $\Re \check{a}_{0}^{[0]-}(\omega)$ (red full), $\Im \check{a}_{0}^{[0]-}(\omega)$ (red dashed),  $\Re \bar{A}^{[0]-}(\omega)$ (black full), $\Im \bar{A}^{[0]-}(\omega)$ (black dashed). Upper right-hand panel: $\Re a_{0}^{[1](M)}(\omega)$ (blue full), $\Im a_{0}^{[1](M)}(\omega)$ (blue dashed),  $\Re \check{a}_{0}^{[1]}(\omega)$ (red full), $\Im \check{a}_{0}^{[1]}(\omega)$ (red dashed),  $\Re \bar{A}^{[1]}(\omega)$ (black full), $\Im \bar{A}^{[1]}(\omega)$ (black dashed). Lower left-hand panel: $\rho^{(M)}(\omega)$ (blue full), $\check{\rho}(\omega)$ (red full) $\bar{\mathcal{R}}(\omega)$ (black full). Lower right-hand panel: $\alpha^{(M)}(\omega)$ (blue full), $\check{\alpha}(\omega)$ (red full) $\bar{\mathcal{A}}(\omega)$ (black full), output fluxes (dashed). $h=0.06~m$,  $w=0.028~m$, $d=0.04~m$,  $\theta^{i}=0^{\circ}$, $M=3$, the constitutive parameters of the foam are those of sect. \ref{gdd}.}
\label{Mgdd 01}
\end{center}
\end{figure}

\begin{figure}[ptb]
\begin{center}
\includegraphics[width=0.75\textwidth]{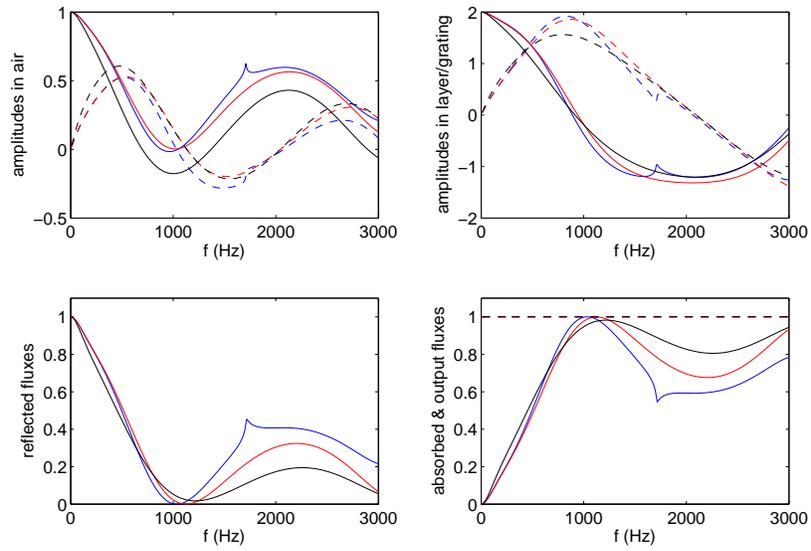}
\caption{Same as fig. \ref{Mgdd 01} except that   $w=0.14~m$, $d=0.2~m$.}
\label{Mgdd 02}
\end{center}
\end{figure}

\begin{figure}[ptb]
\begin{center}
\includegraphics[width=0.75\textwidth]{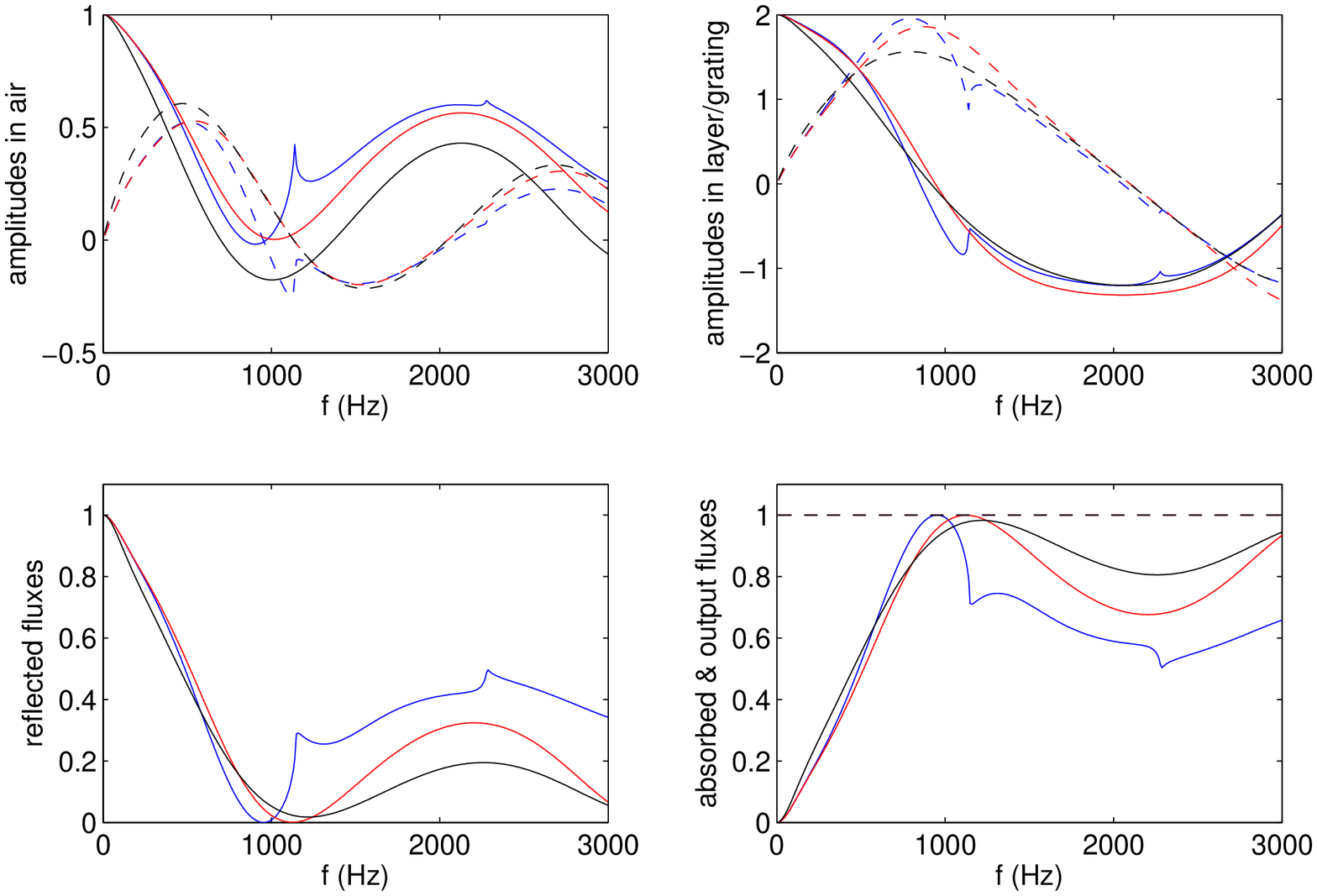}
\caption{Same as fig. \ref{Mgdd 01} except that    $w=0.21~m$, $d=0.3~m$.}
\label{Mgdd 03}
\end{center}
\end{figure}

\begin{figure}[ptb]
\begin{center}
\includegraphics[width=0.75\textwidth]{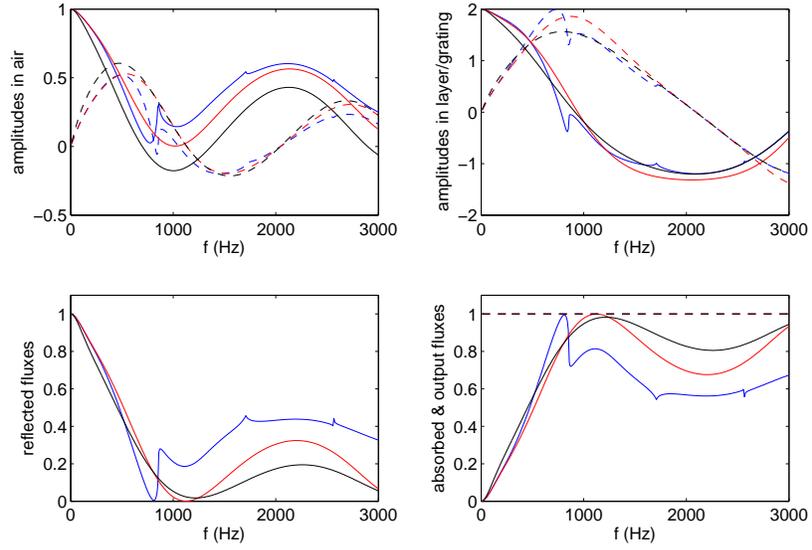}
\caption{Same as fig. \ref{Mgdd 01} except that   $w=0.28~m$, $d=0.4~m$.}
\label{Mgdd 04}
\end{center}
\end{figure}

Figs. \ref{Mgdd 01}-\ref{Mgdd 04} apply to the foam of sect. \ref{gdd}.
\clearpage
\newpage
\subsubsection{$\Phi_{\text{opt}}=0.1$, $h=0.012~m$, and the foam of \cite{glv10}}
\begin{figure}[ht]
\begin{center}
\includegraphics[width=0.75\textwidth]{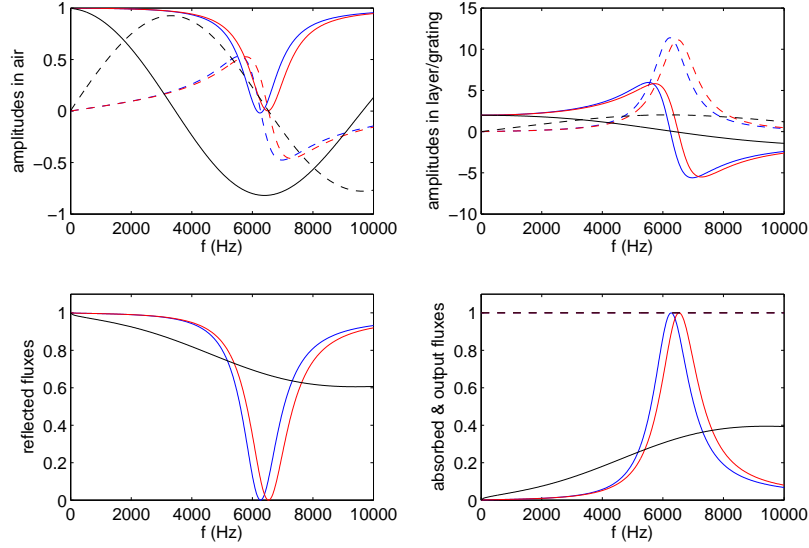}
\caption{Upper left-hand panel: $\Re a_{0}^{[0]-(M)}(\omega)$ (blue full), $\Im a_{0}^{[0]-(M)}(\omega)$ (blue dashed),  $\Re \check{a}_{0}^{[0]-}(\omega)$ (red full), $\Im \check{a}_{0}^{[0]-}(\omega)$ (red dashed),  $\Re \bar{A}^{[0]-}(\omega)$ (black full), $\Im \bar{A}^{[0]-}(\omega)$ (black dashed). Upper right-hand panel: $\Re a_{0}^{[1](M)}(\omega)$ (blue full), $\Im a_{0}^{[1](M)}(\omega)$ (blue dashed),  $\Re \check{a}_{0}^{[1]}(\omega)$ (red full), $\Im \check{a}_{0}^{[1]}(\omega)$ (red dashed),  $\Re \bar{A}^{[1]}(\omega)$ (black full), $\Im \bar{A}^{[1]}(\omega)$ (black dashed). Lower left-hand panel: $\rho^{(M)}(\omega)$ (blue full), $\check{\rho}(\omega)$ (red full) $\bar{\mathcal{R}}(\omega)$ (black full). Lower right-hand panel: $\alpha^{(M)}(\omega)$ (blue full), $\check{\alpha}(\omega)$ (red full) $\bar{\mathcal{A}}(\omega)$ (black full), output fluxes (dashed). $h=0.012~m$,  $w=0.001~m$, $d=0.01~m$,  $\theta^{i}=0^{\circ}$, $M=6$, the constitutive parameters of the foam are those of sect. \ref{glv}.}
\label{Mglv 01}
\end{center}
\end{figure}

\begin{figure}[ptb]
\begin{center}
\includegraphics[width=0.75\textwidth]{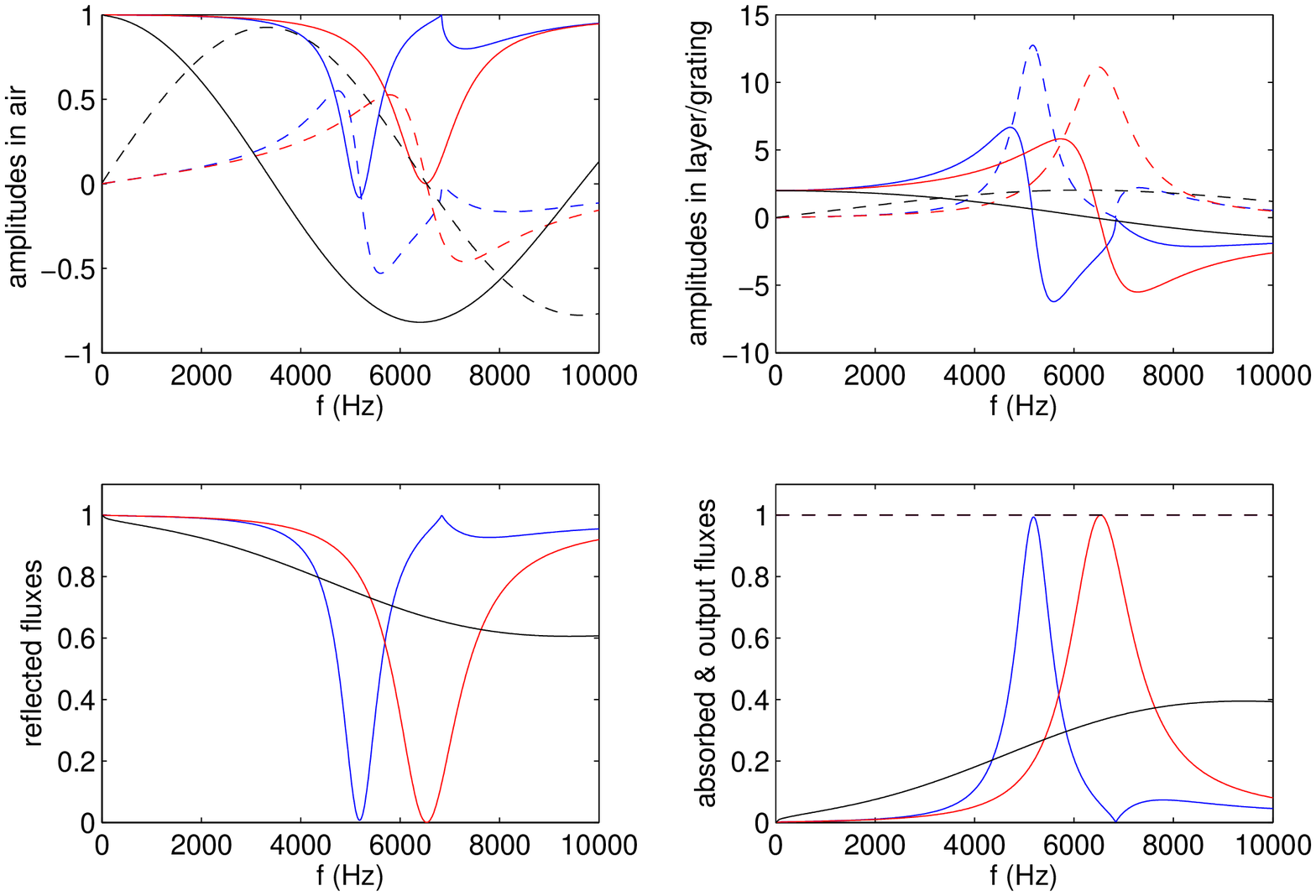}
\caption{Same as fig. \ref{Mglv 01} except that     $w=0.005~m$, $d=0.05~m$.}
\label{Mglv 02}
\end{center}
\end{figure}

\begin{figure}[ht]
\begin{center}
\includegraphics[width=0.75\textwidth]{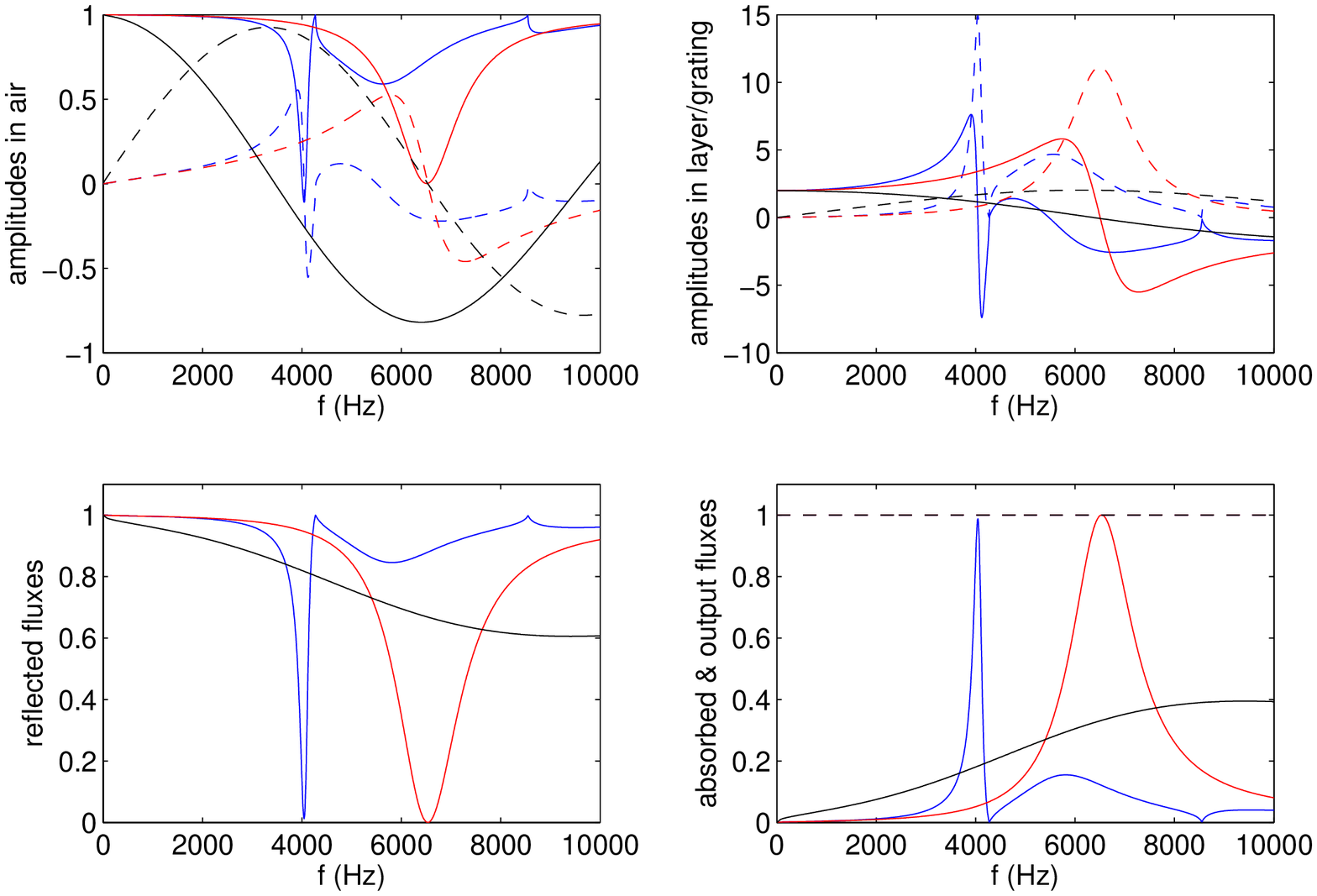}
\caption{Same as fig. \ref{Mglv 01} except that  $w=0.008~m$, $d=0.08~m$.}
\label{Mglv 03}
\end{center}
\end{figure}

\begin{figure}[ptb]
\begin{center}
\includegraphics[width=0.75\textwidth]{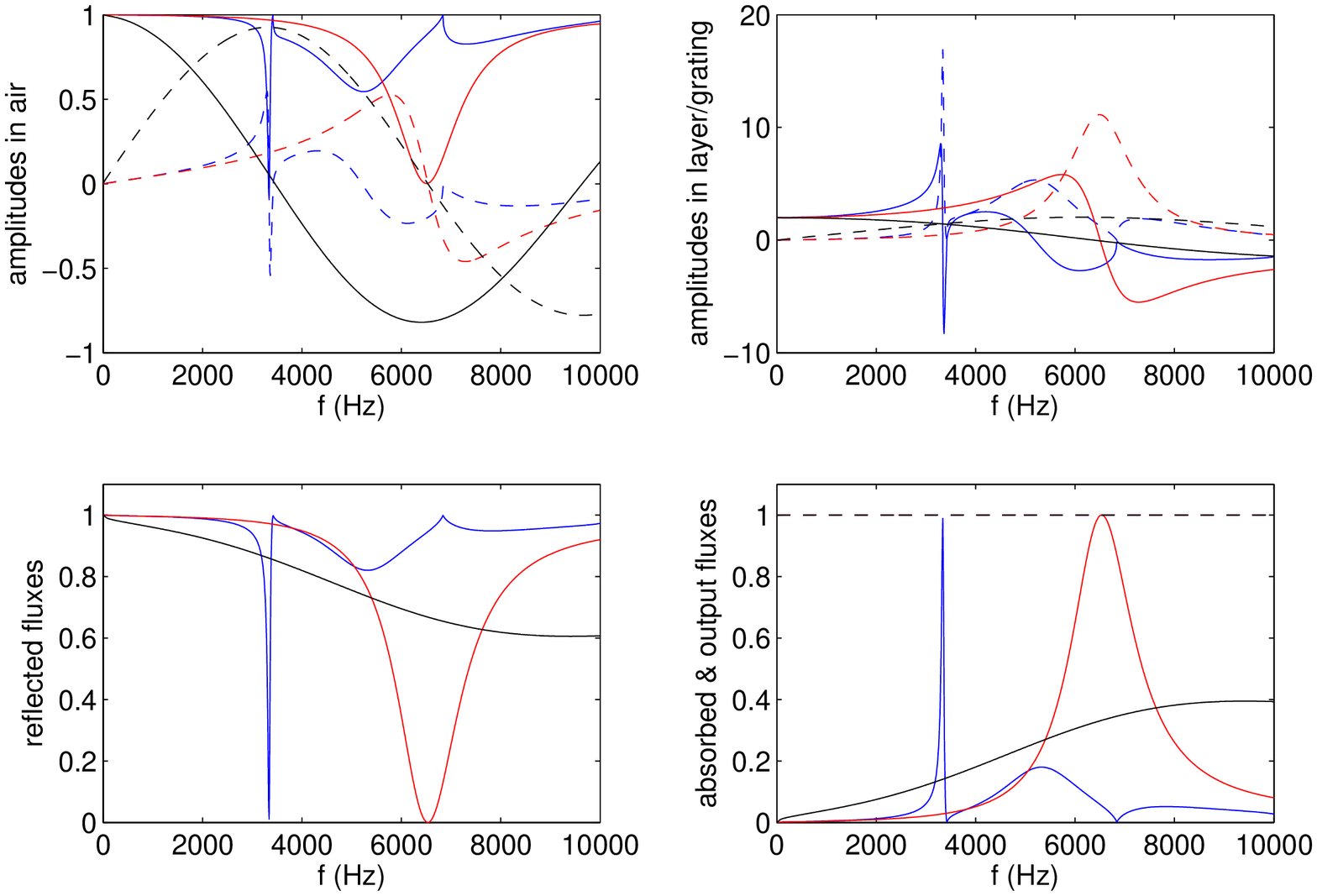}
\caption{Same as fig. \ref{Mglv 01} except that $w=0.010~m$, $d=0.10m$.}
\label{Mglv 04}
\end{center}
\end{figure}

Figs. \ref{Mglv 01}-\ref{Mglv 04} apply to the foam of sect. \ref{glv}.
\clearpage
\newpage

Note that the maximum absorption is nowhere near the $500~Hz$ area, even for the largest $w$ and $d$, and nowhere near the $\sim 700~Hz$ peak in fig. 4 of \cite{glv10}.
\subsubsection{Comments on the 'red shift' of the total absorption peaks}
1. The shift to lower frequencies (i.e., the 'red shift') of the total absorption peak is obtained only for $\alpha^{(M>0)}$ (i.e., neither for $\tilde{\alpha}$ which depends only on the ratio $w/d$, nor for $\bar{\mathcal{A}}$ which does not even depend on $w/d$) by increasing both $w$ and $d$ while maintaining their ratio $w/d$ at the chosen value $\Phi_{\text{opt}}$.\\
2. The bandwidth of substantial absorption near this peak decreases with increasing $w$ and $d$.\\
3. The difference between the grating absorption and the reference layer absorption at the location of the maximum absorption increases with increasing $w$ and $d$ essentially because the absorption of the reference configuration decreases with frequency in the region of maximal grating absorption.\\
4. Substantial very-low frequency absorption cannot be obtained other than in a relatively-small bandwidth and with relatively thick gratings.\\
5. Nothing close to the results in fig. 4 of \cite{glv10} have been obtained for our grating with the foam material of \cite{glv10} .\\
6. The red shift (and Wood anomalies \cite{ms16}, marked by kinks or near-discontinuous behavior in response functions) can probably be explained using the $M=1$ approximation of grating response as is done in \cite{wi18b}.
\clearpage
\newpage
\subsection{Homothetic increase of $h$,$w$ and $d$ for $\Phi_{\text{opt}}=0.4$, $h/d=1$, and the foam of \cite{gdd11}}
\begin{figure}[ht]
\begin{center}
\includegraphics[width=0.75\textwidth]{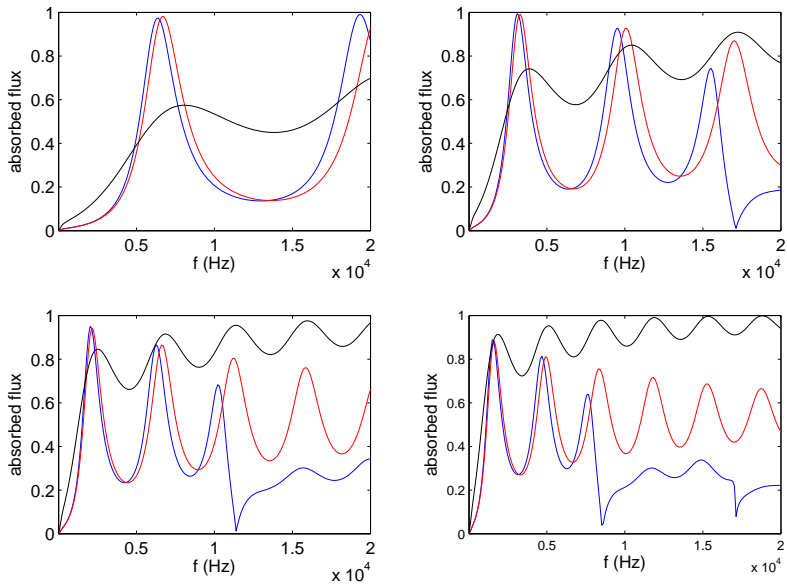}
\caption{ Upper left-hand panel: $\{h,w,d\}=\{0.01~m,0.0025~m,0.01m\}$. Upper right-hand panel: $\{h,w,d\}=\{0.02~m,0.0050~m,0.02m\}$. Lower left-hand panel: $\{h,w,d\}=\{0.03~m,0.0075~m,0.03m\}$. Lower left-hand panel: $\{h,w,d\}=\{0.04~m,0.0100~m,0.04m\}$.}
\label{homot 01}
\end{center}
\end{figure}

In the previous figure, $h$ was kept constant so that the condition of optimality was maintained. Now (fig. \ref{homot 01}) we  augment $h$ while augmenting $w$ and $d$ in the same proportions, so that we are thus increasingly departing from the condition of optimality.
\clearpage
\newpage

This prediction is confirmed in the figure. Moreover, for the largest $h$ we can even arrive at a situation in which the height of the lowest frequency peak is lower than that the height of the lowest frequency reference layer peak, which means that for this choice of parameters, making the layer inhomogeneous is of no use.
\section{Conclusion}
 As mentioned in the Introduction, the present investigation was inspired by what we learned in \cite{gdd11}, and, in particular by the numerical result   exhibited in fig. 3  of this paper concerning the gain in low-frequency absorption enabled by replacing a rigidly-backed foam-filled layer of thickness $h$ by the same layer containing a periodic (along the $x$ axis) distribution of rigid, circular cylinders. Owing to the fact that such a structure is not easy to fabricate, we wondered whether a material interchange of the foam and cylinder components might enable a similar gain of absorption (over that of the reference rigidly-backed foam-filled layer), and if so, would it be possible to explain this gain in a simple manner. To do this, we chose to replace the generic  circular cylinder by a rectangular  cylinder whose height equals $h$ , thus enabling the field representations to be much simpler than for a circular cylinder.

\begin{figure}[ht]
\begin{center}
\includegraphics[width=0.75\textwidth]{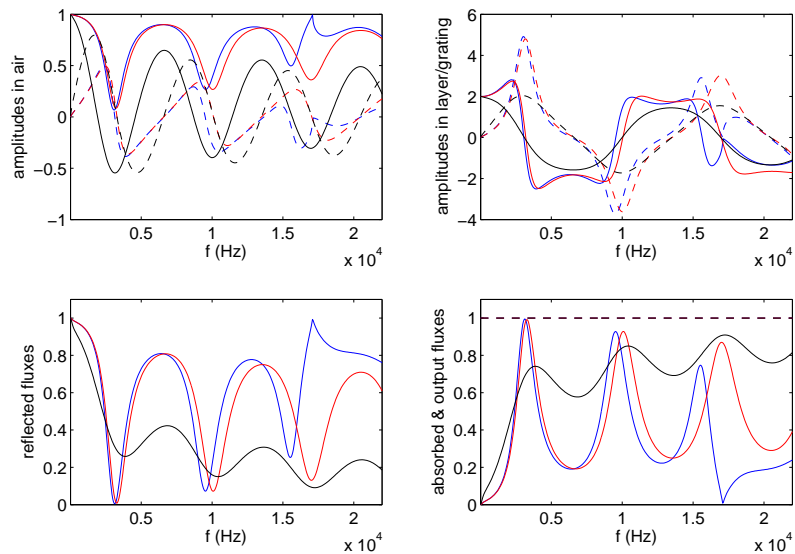}
\caption{Same as fig. \ref{Mgdd 01} except that $h=0.02~m$, $w=0.005~m$, $d=0.02~m$.}
\label{Mgdd 05}
\end{center}
\end{figure}

 Our first numerical trials led to a numerical result depicted   in  fig. \ref{Mgdd 05} concerning the absorption of a grating composed of a periodic distribution of foam-filled rectangular cylinder grooves. This grating configuration is rather close to the one in fig. 3 of \cite{gdd11}, notably by the fact that the thickness of both gratings are the same, i.e., $h=0.02~m$, but, of course, the host medium in the layer is now rigid (instead of being a foam) and the medium in the grooves is now a foam (instead of being rigid).
Despite this difference, something quite similar, notably as concerns the position and height of the reference layer and grating lowest frequency peaks,  to the results in fig. 3 of \cite{gdd11} is observed for our grating with the same foam material as in \cite{gdd11}. The gain of absorption in our lowest-frequency peak (over that of the  lowest-frequency reference layer peak) is obtained for a transverse area of foam material of $10^{-4}~m^{2}$ per period whereas the comparable absorption gain of \cite{gdd11} is obtained for a transverse area of foam material of $2.233\times 10^{-4}~m^{2}$ per period which means that the same absorption gain is obtained in our configuration with less than twice the amount of foam material employed in \cite{gdd11}, even though the two gratings  have the same thickness. Moreover, our grating appears to be structurally more sound and easier to fabricate than the grating of rigid circular cylindrical inclusions within a foam layer.

The main advantage of our grating is that it enables a very simple rigorous analysis of its response to an acoustic wave as well as an even-simpler, mathematically-explicit, approximation of its low-frequency response. The expression of this low-frequency (amplitude $\tilde{A}^{[1]}(\omega)$)  response in the heterogeneous layer region of the rigidly-backed grating configuration turns out to be mathematically identical to the corresponding response in the same region  of a rigidly-backed homogeneous (effective) foam layer configuration of the same thickness as that of the grating provided that the (effective) density  $R^{[1]}$ of the layer is taken to be the density $\rho^{[1]}$ of the foam filler of the grating grooves  divided by the areal filling fraction $w/d$, wherein $w$ is the generic groove width and $d$ the grating period. We showed, by comparison with the rigorous solution, that $\tilde{A}^{[1]}(\omega)$ is a quite accurate approximation of the rigorous amplitude $a_{0}^{[1]}(\omega)$ in the low-frequency region of interest and for incident angles that do not exceed $\sim 40^{\circ}$. Moreover we showed that the approximation of the absorbed flux $\tilde{\mathcal{A}}$ is expressed as the product of a  term involving $\|\tilde{A}^{[1]}(\omega)\|^{2}$ (that depends on $w/d$) with a trigonometric term $E(\omega)$ that does not depend  on $w/d$ but rather on the thickness $h$. It is not easy to explain theoretically (for a given $h$) the position and height of  the lowest-frequency absorption peak produced by our grating because of the dispersive nature of the foam material, but a numerical study has enabled to establish the principal tendencies. Above all, the expression for $\tilde{A}^{[1]}(\omega)$ can differentiated with respect to $\Phi=w/d$ in order to find the optimal $\Phi=\Phi_{opt}$ (i.e., in the sense of maximizing the absorption), and from this  even find the frequency at which total absorption can be expected to occur. Subsequently, we showed theoretically that the absorption produced by the so-optimized grating is always equal to or greater than that of the homogeneous reference layer (also of thickness $h$) filled with the same foam material as that of the filler material in the grooves of the grating.

As mentioned in the Introduction, it can also useful to  lower the frequency of occurrence of the lowest-frequency absorption peak (such as in the design of anechoic chambers). This can be done by increasing $h$, but usually results in a lowering of height of this peak. We thus decided to find out if another strategy might lead to the desired result. With the approximate response as a starting point, and a $\Phi_{opt}$ obtained from this response as an invariant (i.e., maintaining $w/d$ constant), we chose to increase $w$ and $d$ at the same rate  and, by so doing, depart from the quasi-static situation which previously enabled the simple approximation of response. Thus, we were obliged to resort to a numerical exploitation of our rigorous solution which revealed the interesting fact that increasing $w$ and $d$, while maintaining $w/d$ at the value $\Phi_{opt}$ can enable the lowering of the frequency of occurrence of the first low-frequency absorption  peak without affecting its height (which is maintained at the total absorption level). Unfortunately, this is obtained at the expense of narrowing the bandwidth of large absorption around this peak.

Among the perspectives of this work we may cite;\\
1. The $M=1$ approximation of our 2D grating response will probably enable the explanation (in the manner of  \cite{wi18b}) of Wood's anomalies and the red shift of the total absorption peak;\\
2. It should be rather easy to extend, as in \cite{bd05}, this type of rigorous and approximate analysis to a 2D grating  composed of a periodic (in the $x$ and $y$ directions) array of foam filled box-shaped troughs, the interesting question being whether this structure will enable even larger absorption at even lower frequencies.\\
3. It should be rather easy to extend (in the manner of \cite{ak76}) this type of rigorous and approximate analysis to a 1D grating with structured (e.g., multi-stepped) grooves \cite{pt75}, the interesting question again being whether this may enhance the sought-for effects;\\
5. It might be possible to enhance the absorption at very low frequencies ($<500~Hz$) by appealing to  better filler materials \cite{ac10,at18}, and the present analysis could be applied to such configurations without fundamental changes ((i.e., only the effective parameters $\rho_{e}(\omega)$ and $c_{e}(\omega)$ change from their values given in sects. \ref{gdd} and \ref{glv}).
%

\end{document}